\documentclass[preprint,elsarticle-num]{elsarticle}
\usepackage[utf8]{inputenc}
\usepackage{listings}

\usepackage[author={Ralf}]{pdfcomment}
\usepackage{color,xcolor}
\usepackage{cancel}
\usepackage{ulem}
\newcommand{\indicate}[1]{\textbf{\color{green}#1}}

\usepackage{amssymb}
\usepackage{amsmath}
\usepackage{graphicx}

\usepackage{hyperref}
\usepackage{url}
\usepackage{verbatim}
\usepackage{siunitx}

\usepackage{float}
\usepackage[numbers]{natbib}

\usepackage{JournalNames}

\usepackage[a4paper, margin=4cm]{geometry}

\newcommand{\ISRFGalprop}{GALPROP ISRF }
\newcommand{\ISRFPicard}{PICARD ISRF }

\newcommand{\lt}{\ensuremath <}

\begin{document}

\begin{frontmatter}
\title{The consequence of a new ISRF model of the Milky Way on predictions for diffuse gamma-ray emission}
\author[ibk_ap]{F.~Niederwanger\corref{cor_auth}}
\ead{Felix.Niederwanger@uibk.ac.at}
\author[ibk_ap]{O.~Reimer}
\author[ibk_ap]{R.~Kissmann}
\author[MPIE]{A.~W.~Strong}
\author[pop_aff1,pop_aff2]{\\C.C.~Popescu}
\author[tuffs_aff]{R. Tuffs}

\address[ibk_ap]{Institut f\"ur Astro- und Teilchenphysik,	  Leopold-Franzens-Universit\"at Innsbruck, A-6020 Innsbruck, Austria}

\address[MPIE]{Max Planck Institut f\"ur extraterrestrische Physik, Postfach 1312, D-85741 Garching, Germany}

\address[pop_aff1]{Jeremiah Horrocks Institute, University of Central Lancashire, PR1 2HE, Preston , UK}
\address[pop_aff2]{The Astronomical Institute of the Romanian Academy, Str. Cutitul de Argint 5, Bucharest, Romania}
\address[tuffs_aff]{Astrophysics Department, Max Planck Institut für Kernphysik}

\cortext[cor_auth]{Corresponding author}

\begin{abstract}
We investigate the impact of the recently published ISRF model of the Milky Way by
Popescu et al. on the CR 
electrons and positrons in the context of cosmic-ray transport modelling.
We also study predictions for diffuse Galactic gamma radiation and underline
the importance of the increased ISRF intensities.
We use the PICARD code for solving the CR transport equation to obtain predictions based on Galactic cosmic-ray electron fluxes.
We show that the new ISRF yields gamma-ray intensity increases, most particular
at the Galactic Center and in the Galactic Plane. The impact is largest at energies around \SI{220}{GeV}.
\end{abstract}


\end{frontmatter}


\section{Introduction}
The gamma-ray sky above 100 MeV is dominated by diffuse Galactic emission (DGE) \citep{2012ApJ...750....3A},
which is produced by cosmic rays (CR) that are interacting with the Galactic gas and the interstellar radiation field (ISRF). 
The DGE is an important source for background radiation of gamma-ray observations within the Galactic Plane and the Galactic halo.
Since cosmic rays are deflected and fragment/decay while they propagate through the Galaxy, diffuse gamma-ray emission provides a suitable tool to study their origin and propagation properties \citep{annurev.nucl.57.090506.123011}.

There is indication for an excess in the diffuse gamma-ray emission above \SI{10}{GeV} in the Galactic Center and in the Galactic Plane
with respect to widely used models \citep{2012ApJ...750....3A,HOOPER2011412,0004-637X-753-1-41,PhysRevD.84.123005} based on the GALPROP code\footnote{\url{https://galprop.stanford.edu/}} 
\citep{0067-0049-223-2-26,PhysRevD.84.123005}.
One of the applications are the 
models of the Fermi-LAT collaboration \citep{0067-0049-223-2-26}, which have been tuned and/or use re-normalized CR fluxes in order to reproduce the local observational data and are able to accurately describe a wide number of experiments.
Alternative models propose different solutions to the problem, e.g. by introducing galactocentric rigidity dependence of the diffusion 
coefficient (DRAGON Code\footnote{\url{http://dragonproject.org}} \citep{gaggero2016gamma}).
Charged cosmic-rays can only be directly measured at Earth, and especially high-energy electrons do not propagate far from their sources,
thus making it difficult to deduce the CR distribution in the inner Galaxy, where the diffuse emission is expected to reach its maximum \citep{0004-637X-819-1-44}.
Additionally, different intensities and spectral shapes of the ISRF components consequently affect gamma-ray predictions from regions like the Galactic Center.

Prediction for the ISRFs of the Milky Way have been provided by the commonly used GALPROP model \citep{2006ApJ...648L..29P,2006ApJ...640L.155M,2008ApJ...682..400P}, based on models for the stellar emissivity and dust distributions derived from star counts and gas measurements, which are known to have limitations.
Recently, a new model of the stellar and dust radiation fields
has been released \citep{2017MNRAS.470.2539P}, which matches the relevant data from the Milky Way.
We show predictions of a numerical CR propagation model for the gamma-ray emission using the new ISRF model.

\section{The interstellar radiation field model of the milky way}
\label{sc:ISRF}

The new ISRF model of the Milky Way \citep{2017MNRAS.470.2539P} was derived
using realistic radiative transfer models, that were successfully used to explain the spectral energy distribution (SED) of individual nearby galaxies \citep{2000A&A...362..138P,2011A&A...527A.109P,2004A&A...414...45P},
as well as for the statistical behaviour of large samples 
(e.g. \citep{2016MNRAS.461..458D,2008ApJ...678L.101D,2007MNRAS.379.1022D,2012MNRAS.427.3244D,2017AJ....153..111G,2013ApJ...766...59G,2013A&A...553A..80P,2013A&A...557A.137P,2014MNRAS.441.1340V}).
When applied to the Milky Way, these radiative transfer models were adapted to the internal view of the Galaxy and to constraints imposed by our edge-on view of the Galaxy from the position of the Sun.
The spatial and spectral distribution of the ISRF, together with the stellar and dust distributions were derived in a self-consistent way by iteratively comparing models with 
observations from COBE, IRAS, and PLANCK.
The model is thus consistent with imaging data from the near-infrared to the sub-millimeter regime. 
The ISRF consists of two components: stellar-, and dust-emission light. 
We complement this with the cosmic microwave background according to Planck's law \citep{RybickiLightman..RadProcesses}.

\begin{figure}
	\includegraphics[width=\textwidth]{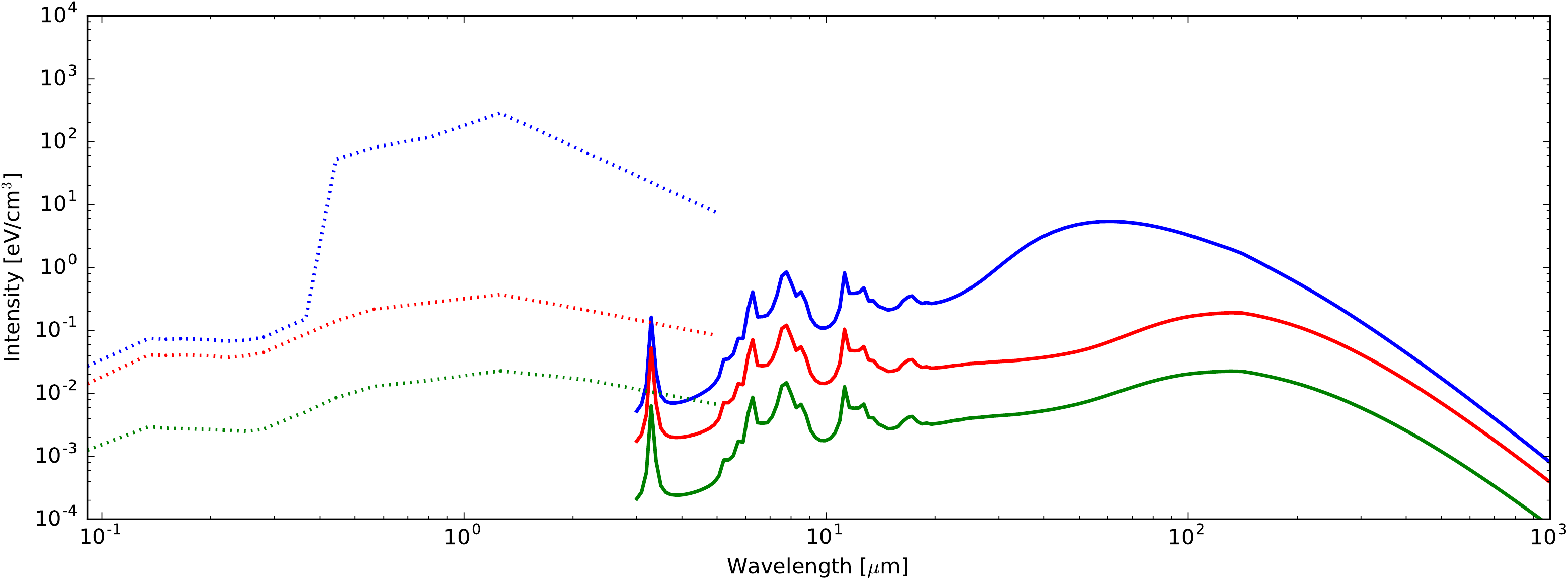}
	\caption{Spectrum of the ISRF components (Starlight is dotted and dust-emitted light is solid) at the Galactic Center (blue), Earth (red) and at $R = 15$ kpc, $z = 0$ kpc (green)}
	\label{fig:ISRFComponents}
\end{figure}

The ISRF model is defined on a cylindrical coordinate system $(r,z)$ at different wavelengths $\lambda$:
$S = S(r, z, \lambda)$.
The frequency binning allows for accurate inclusion of prominent lines and has higher resolution than previous models.
The new ISRF model comprises $135$ frequency bins, with $15$ bins representing the starlight and $120$ bins the dust-emitted light component.
The spatial grid spans $22$ points in the $r$ and $z$ direction, respectively ranging from $0$ to $24$ kpc
in radial direction and $0$ to $10$ kpc in vertical direction.
Those points are on a logarithmic scale,
leading to a high-resolution description near the Galactic Center.
The ISRF intensity decreases notably with increasing distance from the Galactic
Center, with intensities changing by nearly two orders of magnitude for the starlight component between
the Galactic Center and Earth. Figure \ref{fig:ISRFComponents} shows the starlight and dust emission component of the ISRF model at different distances from the Galactic Center.

The starlight component in the ISRF model in \citet{2017MNRAS.470.2539P} reaches up to \SI{5}{\micro\meter}.
At this wavelength, the starlight intensity is still approximately an order of magnitude larger than the dust-emitted light.
We therefore extend the starlight component to longer wavelengths by using a simple power-law fit on the last two available wavelength bins.

The dust emission component ranges from \SI{3}{\micro\meter} to \SI{1}{\milli\meter}. For shorter and longer
wavelengths we use extrapolation on the last two available datapoints. 
There, however, the dust component is already subdominant to the starlight or CMB emission.

\subsection{The ISRF model in PICARD}

\begin{figure}
	\includegraphics[width=\textwidth]{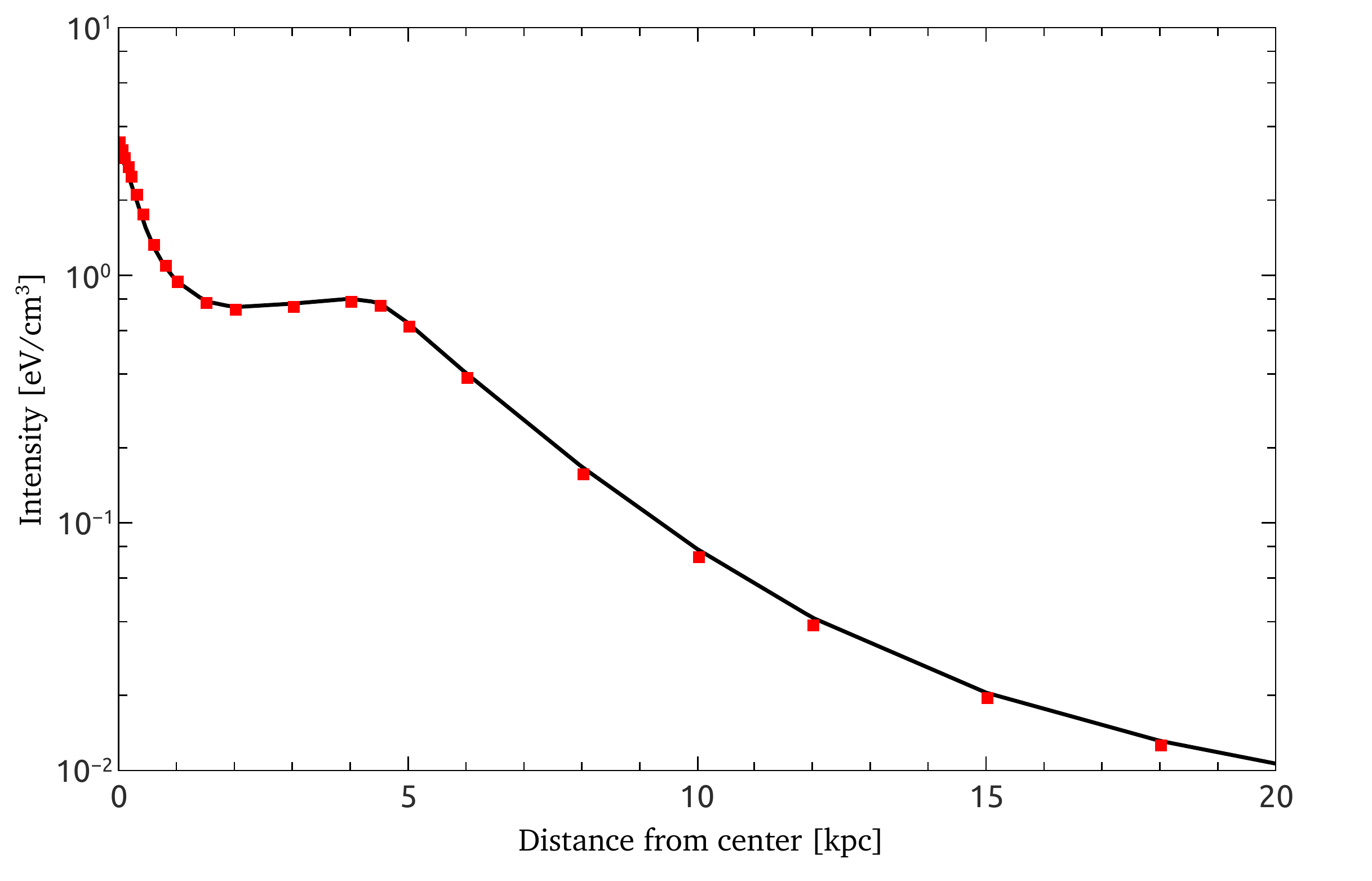}
	\caption{Radial profiles of the Popescu et al. ISRF model (red dots) and corresponding intensity mapped onto the equidistant PICARD grid (black solid line) at \SI{100}{\micro\meter}.}
	\label{fig:ISRFMapping}
\end{figure}

We map the new ISRF model onto the equidistant grid used by
PICARD (see section \ref{sc:CRModel}) via logarithmic interpolation.
As an example Figure \ref{fig:ISRFMapping} shows the ISRF model and the mapped ISRF in PICARD at a wavelength of \SI{100}{\micro\meter}.

Henceforth we refer to the Popescu et al. ISRF model \citep{2017MNRAS.470.2539P} that has been mapped to the PICARD grid as the PICARD ISRF model.

\subsection{Radial and vertical profiles}

Because of its axisymmetric nature, vertical and radial profiles of the ISRF intensity are the most natural 
way to discuss the spatial dependence of the \ISRFPicard model.

\begin{figure}
	\includegraphics[width=\textwidth]{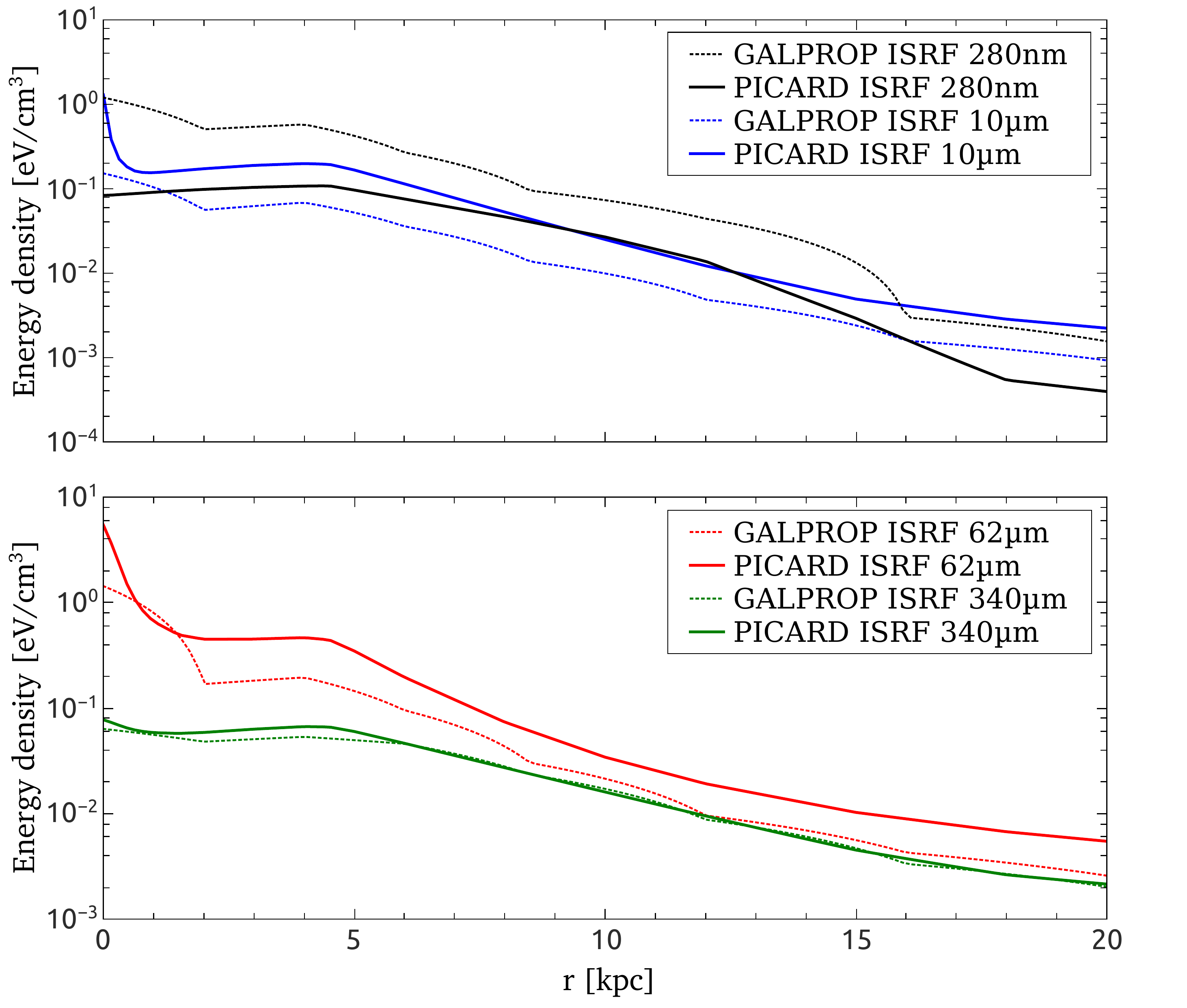}
	\caption{Radial profiles at $z = 0\,$kpc of the two ISRF models investigated in this study at different wavelengths. Solid is the \ISRFPicard model, dotted the \ISRFGalprop model.
	}
	\label{fig:ISRFRadialProfiles}
\end{figure}

\begin{figure}
	\includegraphics[width=\textwidth]{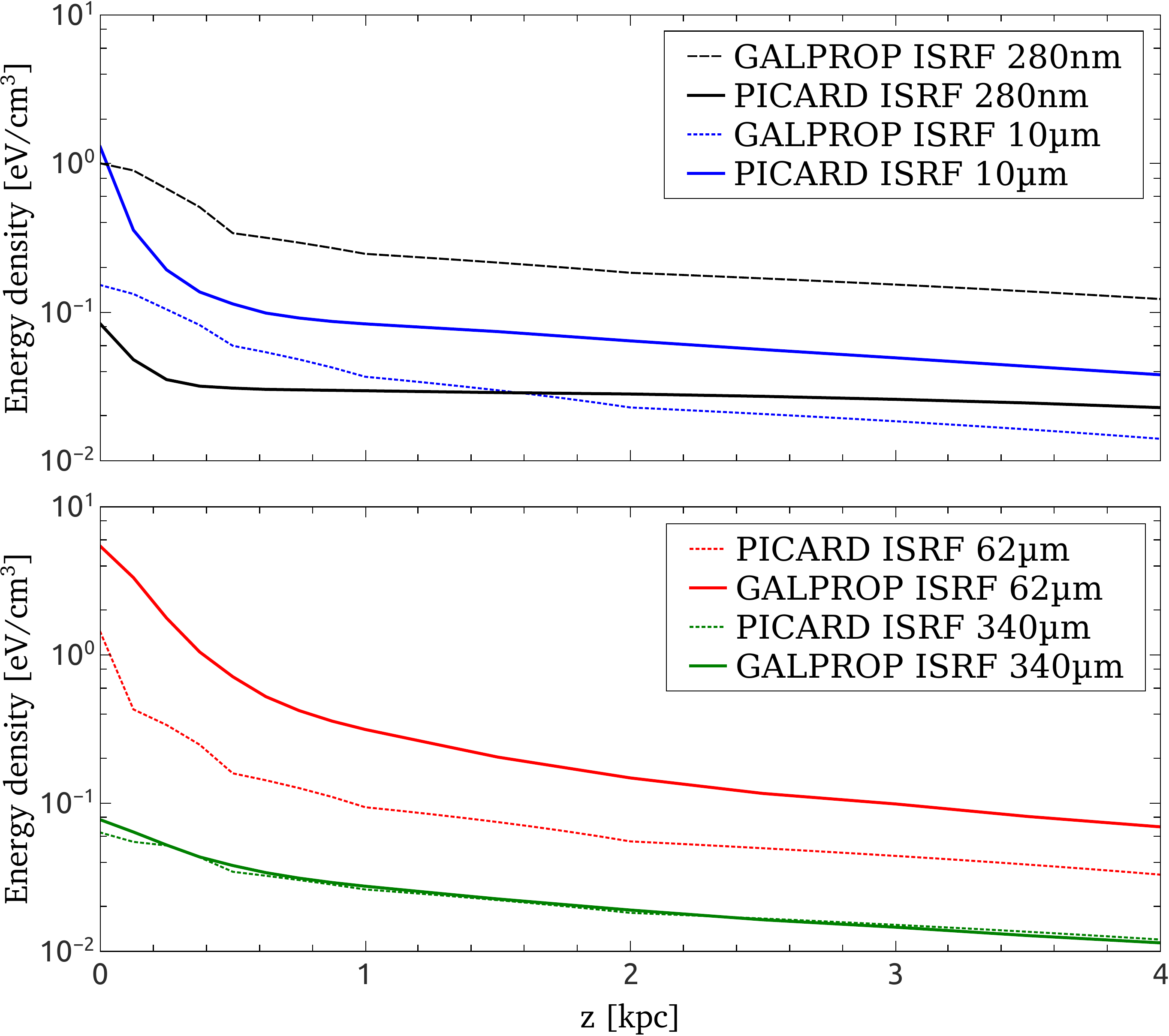}
	\caption{Same as Fig. \ref{fig:ISRFRadialProfiles}, but for the vertical profiles.}
	\label{fig:ISRFAxialProfiles}
\end{figure}

Figure \ref{fig:ISRFRadialProfiles} shows radial profiles whereas Figure \ref{fig:ISRFAxialProfiles} shows
vertical profiles for the \ISRFGalprop model and the \ISRFPicard model at selected wavelengths, respectively.

The intensity gradients, both in the radial and the vertical direction, are different towards the Galactic Center and the Galactic Plane respectively. For the inner \SI{1}{kpc} the radial profiles show an order-of-magnitude general intensity increase for the starlight- and dust-dominated wavelength regimes and similar intensities towards the 
CMB regime.

Between \SI{2}{kpc} and \SI{5}{kpc} the radial profile is nearly flat. 
Similar to the radial profile, the vertical profile shows  an increase in the inner \SI{0.5}{kpc} up to \SI{1}{kpc}, depending on the wavelength, and a flatter profile at a height larger than \SI{1}{kpc}.
In general, the PICARD ISRF has higher intensities towards the Galactic Plane than the GALPROP ISRF in the dust dominated regime.

\subsection{Spectral comparison with previous model}

\begin{figure}
	\includegraphics[width=0.99\textwidth]{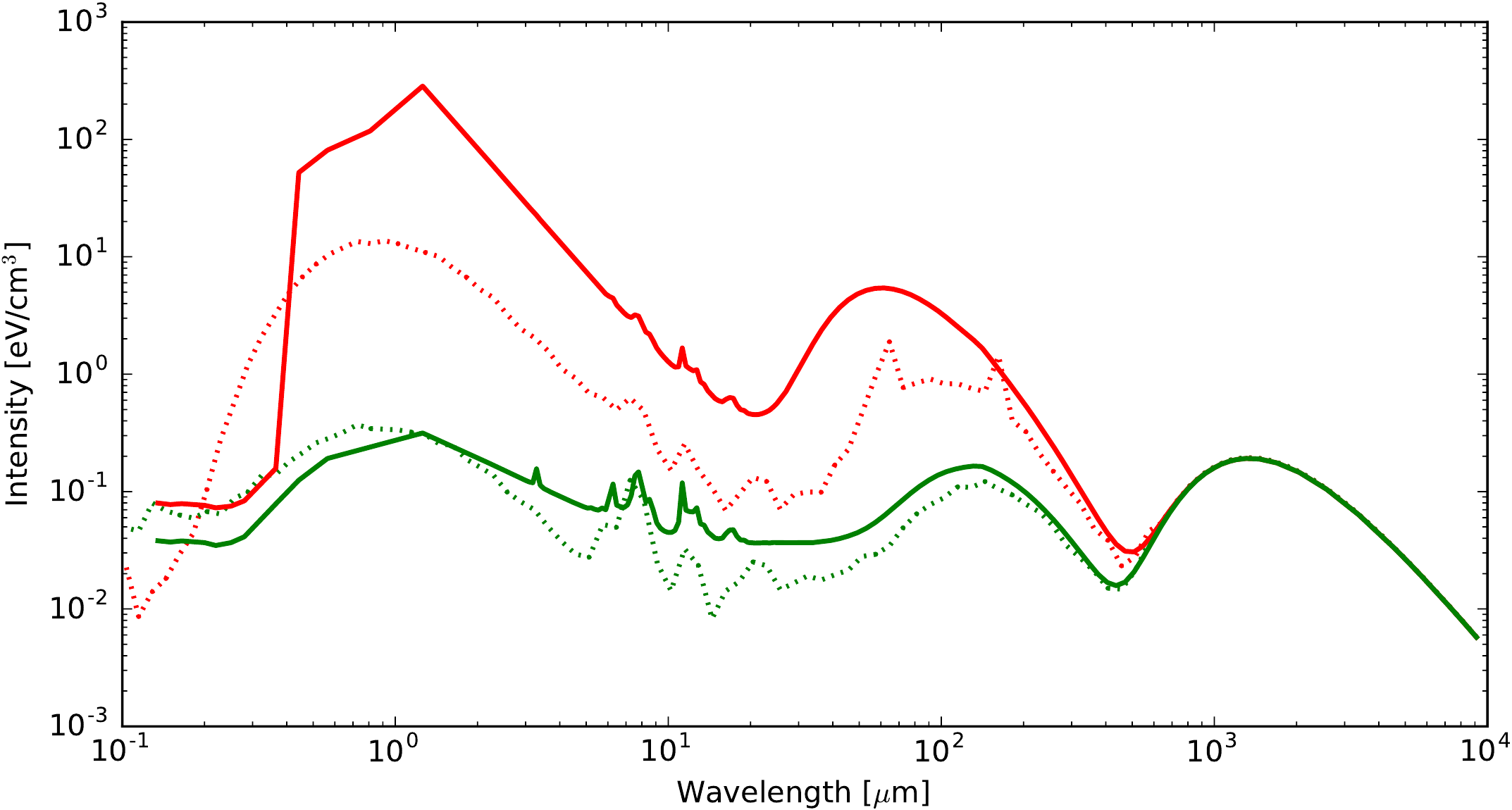}
	\caption{SED for the two ISRF models at the Galactic Center (red, upper) and Earth (green, lower), respectively. Solid line for PICARD ISRF, dotted line for GALPROP ISRF.}
	\label{fig:ISRFSEDDistribution}
\end{figure}

Figure \ref{fig:ISRFSEDDistribution} shows the spectral energy distribution of both ISRF models at Earth and at the Galactic Center.
In contrast to the \ISRFGalprop model, the new ISRF model has three distinct new features:
A substantial intensity increase towards the Galactic Center, 
a generally larger intensity in the dust emission regime, and below \SI{90}{\nano\meter} an overall reduced intensity in the starlight regime.

The ISRF-intensity increase in the Galactic Center dominates 
for wavelengths above \SI{0.4}{\micro\meter} and is limited to the inner \SI{0.5}{kpc} of
the Galaxy. At \SI{0.5}{kpc} the intensities of the two ISRF models are comparable for
wavelengths between \SI{1}{\micro\meter} and \SI{10}{\micro\meter}.

In the dust-dominated regime, the intensity increase is most obvious
between \SI{3}{\micro\meter} and \SI{200}{\micro\meter}. 
At Earth the effect amounts to between a factor of $1.5$ at \SI{3}{\micro\meter} and a factor of $3$ at \SI{30}{\micro\meter}.
At longer wavelengths ($>$ \SI{200}{\micro\meter}) the two ISRF models have comparable intensities.

In the starlight regime, below \SI{90}{\micro\meter} for Earth and \SI{2}{\micro\meter} for the Galactic Center, the intensity of the new ISRF model is smaller compared to the
\ISRFGalprop model.

Apart from this, the spectral resolution of the new ISRF is larger, especially in the dust dominated regime. For example the diffuse interstellar band
features \SI{10}{\micro\meter} are captured more clearly than in previous models used in CR propagation codes with practical no visible consequences for the gamma-ray emission (See section \ref{sc:CRModel}). Reducing the spectral resolution to the one used by GALPROP does not lead to any obvious changes for the gamma-ray emission.
Both ISRF models converge with increasing wavelength and are identical in the CMB.

Here we note that the amplification of the NIR radiation fields at the Galactic Center
in \cite{2017MNRAS.470.2539P} strongly depends on the
assumed optical constants of the dust, which were taken to be the canonical values from
\cite{2001ApJ...548..296W}. These dust absorptivies and
emissivities, although empirically anchored at laboratory measurements
in the UV/optical/NIR/FIR range, are quiet uncertain in the submm regime. If the submm
grain emission efficiencies were higher than predictions of the WD01 model \citep{2001ApJ...548..296W}, this would be an overestimation of the
UV-optical depth of the Galaxy.
Since the amplitude and spectrum
of the stellar emission in the NIR was derived in \cite{2017MNRAS.470.2539P} by fitting
the NIR imaging of the directly observed stellar light, the radiation fields depend sensitively on the assumed dust properties. However, the NIR radiation fields are not 
important contributors to the diffuse gamma-ray emission. On the other hand, the FIR/submm
ISRFs, which are the dominant source of seed photons for the IC emission, are highly
robustly determined, since they are directly mapped into the FIR emissivities.

Apart from the NIR regime, the \citet{2017MNRAS.470.2539P} model for the UV/optical and MIR/FIR/submm energy densities is invariant to the choice of the dust model.


\section{Cosmic-ray propagation model}
\label{sc:CRModel}

We use the PICARD code \citep{CRISM2014...Picard...Kissmann} for numerical modelling of the CR 
propagation.
Our CR propagation model cites the propagation parameters $\mathrm{^SY^z4^R20^T150^C5}$\footnote{Nomenclature taken from \citep{2012ApJ...750....3A}, The radial extent is $R = $ \SI{20}{kpc}, the height $z = $ \SI{4}{kpc}, $T_S = $\SI{150}{K} is a assumed unified hydrogen spin temperature, and the $E(B-V)$ cut $ = 5 \cdot 10^{20} cm^{-2}mag^{-1}$.
} from \citep{2012ApJ...750....3A} using an axisymmetric CR source distribution
based on pulsars  from \cite{2004A&A...422..545Y}.
The electron injection spectrum is a broken power law with spectral index of $1.6$ ($\le$ GeV), $2.42$ ($\sim2$ GeV to $2$ TeV) and $4$ ($>2$ TeV).

We compare with the publicly available GALPROP code v54 and the thereby provided ISRF model referred further on as \ISRFGalprop model \citep{1998ApJ...509..212S, 2006ApJ...648L..29P}. 

We perform simulations for both ISRF models using $257$x$257$x$65$ grid points in space, covering an extent from $-20$ to $20$ kpc in $x$ 
and $y$ direction and $-4$ to $4$ kpc in $z$ direction,
and $127$ points in momentum space ranging from \SI{10}{MeV} to \SI{1}{PeV} \citep{2015APh....70...39K}

Predictions of the gamma-ray emissivity include Bremsstrahlung, neutral pion decay, and IC emission
for an energy range between \SI{50}{MeV} and \SI{10}{TeV}.


\section{Application of the ISRF}
\label{sc:Application}

In this section we investigate the effect of the new ISRF on cosmic ray propagation
(Sections \ref{sc:ApplicationEnergyLossRates} and \ref{sc:CRFluxes})
and the subsequent prediction of gamma-ray emission (Section \ref{sc:GammaRays}), based on the transport model discussed in \ref{sc:CRModel}.



\subsection{Energy loss rates}
\label{sc:ApplicationEnergyLossRates}

\begin{figure}
	
	\setlength{\unitlength}{0.001\textwidth}
	\begin{picture}(881,596)(-50,0)
	\put(-50,230){\rotatebox{90}{$E^2\left( dE/dt \right)$ [MeV$^2$ eV/s]}}
	\includegraphics[width=\textwidth]{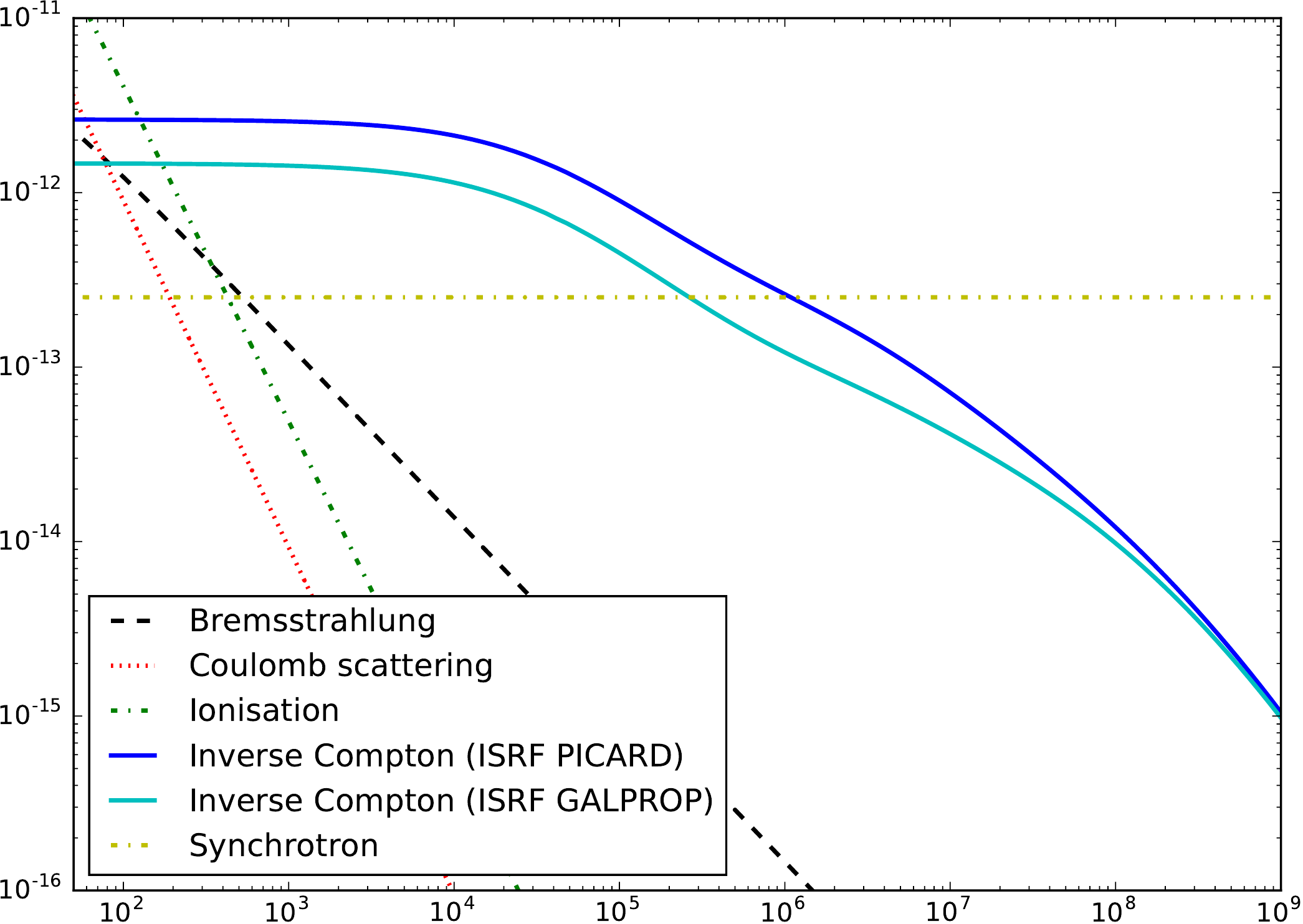}
	\end{picture}
	\begin{picture}(602,596)(-50,-50)
	\put(-50,140){\rotatebox{90}{$E^2\left( dE/dt \right)$ [MeV$^2$ eV/s]}}
	\put(420,-40){$E$ [MeV]}
	\includegraphics[width=\textwidth]{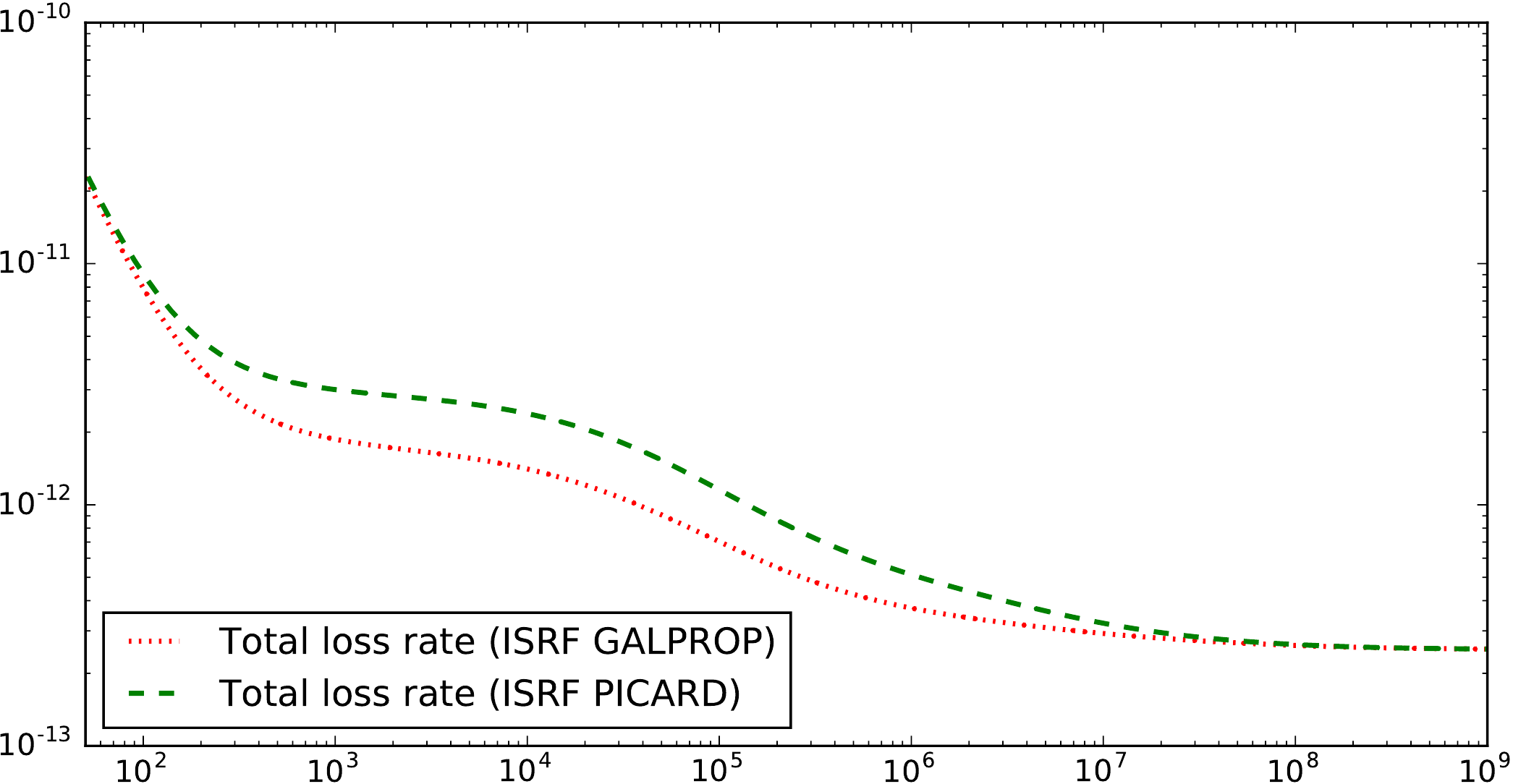}
	\end{picture}
	
	\caption{Individual (upper panel) and total (lower panel) energy loss rates for electron at the Galactic Center for both ISRF models.}
	\label{fig:ElectronLossRates}
\end{figure}

In the following we consider Bremsstrahlung, Coulomb scattering, ionisation losses, IC and synchrotron radiation.
Figure \ref{fig:ElectronLossRates} shows the individual and total energy loss rates for
electrons at the Galactic Center. The order-of-magnitude intensity increase in the \ISRFPicard compared to the \ISRFGalprop translates
to a considerable increase of IC losses up to roughly \SI{100}{TeV}.
At higher energies the Klein-Nishina
cross section equalizes the IC loss rate between the two ISRF models, with synchrotron
already being the dominant loss process.
IC losses dominate the total energy loss rate, up to
$\sim$ \SI{300}{GeV} for the \ISRFGalprop model or up to \SI{1.5}{TeV} for
the \ISRFPicard model.

As discussed in \citep{2017MNRAS.470.2539P}, the most relevant seed photons for IC gamma-rays in the GeV to TeV regime originate from the dust-dominated regime
because higher energy seed photons from the starlight regime are already in the relativistic Klein-Nishina regime.


\begin{figure}[h]
	\includegraphics[width=\textwidth]{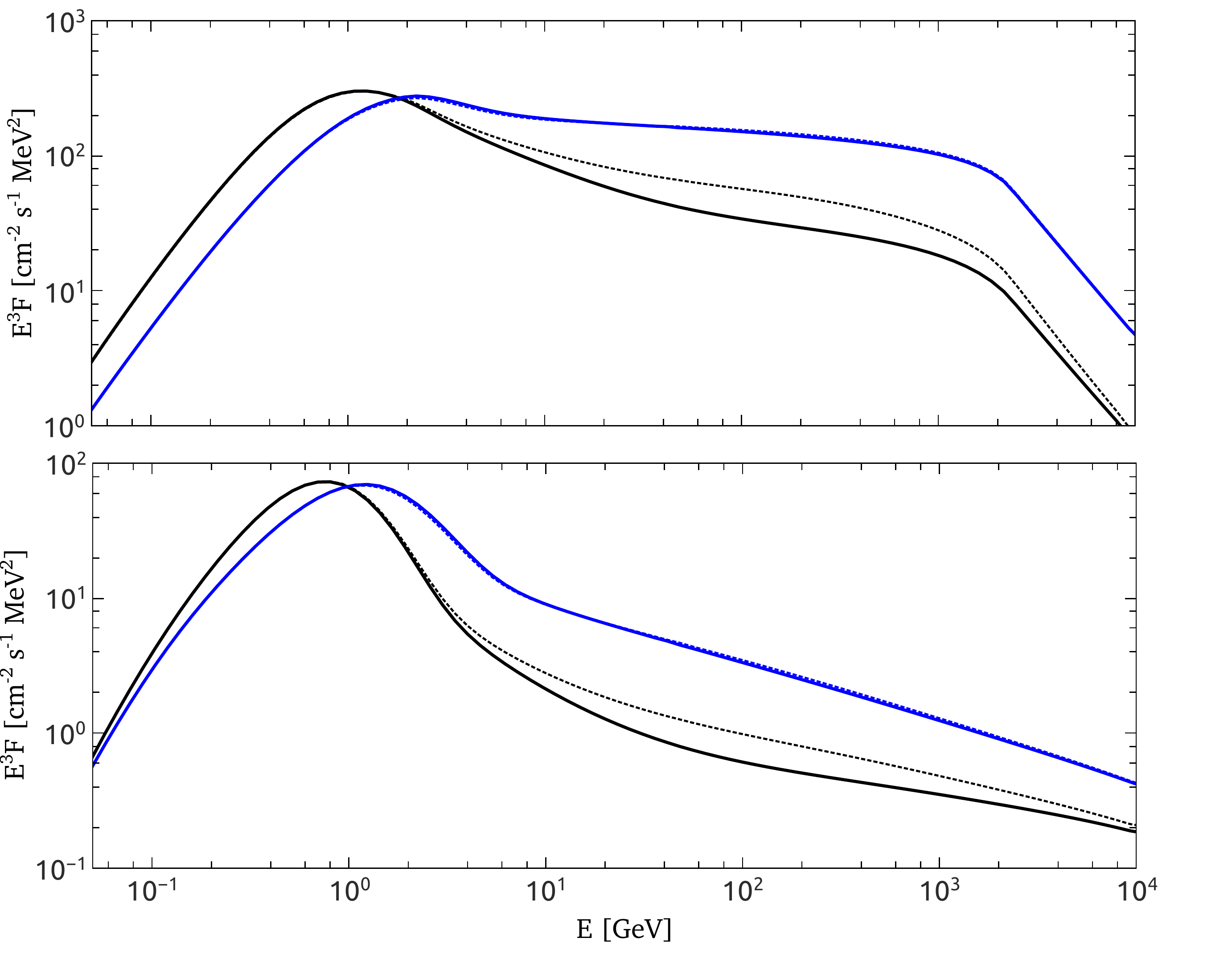}
	\caption{Upper panel: Electron spectra at Earth (blue) and the Galactic Center (black) for the \ISRFGalprop model (dotted) and the \ISRFPicard model (solid), lower panel: likewise for positrons}
	\label{fig:LeptonSpectra}
\end{figure}

\subsection{Cosmic-ray fluxes}
\label{sc:CRFluxes}
Figure \ref{fig:LeptonSpectra} shows results from modelling electron
and positron fluxes at the Galactic Center and at Earth using the two different ISRF models.
The electron spectrum fits a simple power-law from \SI{10}{GeV} to \SI{500}{GeV} and above \SI{3}{TeV}, with the breaks stemming from the electron injection spectrum.
The electron and positron spectra at Earth are very similar for both ISRF models. 
The spectral index of the electron flux for the energy range between $10$ and \SI{500}{GeV} is $\approx 3.08$ for the \ISRFGalprop and $\approx 3.10$ for the \ISRFPicard model. For comparison,  PAMELA data indicate a spectral index of $3.18 \pm 0.05$ above \SI{30}{GeV} \citep{PhysRevLett.106.201101},
Fermi-LAT of $3.04$ between \SI{20}{GeV} and \SI{1}{TeV}
\citep{abdo2009measurement} and H.E.S.S. data an index of $3.05 \pm 0.02$ with a softening
 at $2.1 \pm 0.3$ TeV\citep{aharonian2008energy}.
DAMPE reports a spectral index of $3.1$ for electrons above $>$ \SI{55}{GeV}, 
with a spectral break at $\approx$ \SI{0.9}{TeV} to a spectral index of $3.9$ up to \SI{2.63}{TeV}\citep{ambrosi2017direct} and
CALET reports a single electron power law in the range from \SI{10}{GeV} to \SI{3}{TeV}
with a spectral index of $-3.152 \pm 0.016$\citep{PhysRevLett.119.181101}.

In our models, at higher energies, the electron spectrum softens to $4.70 \pm 0.05$ between \SI{3}{TeV} and \SI{10}{TeV} for both ISRF models.

In the Galactic Center, the use of the new ISRF leads to a considerable softening of the electron spectrum above \SI{3}{GeV}. The corresponding decrease in the electron flux between \SI{3}{GeV} to \SI{10}{TeV} is at most 40\% at \SI{150}{GeV}.



Given the underrepresentation of positrons against electrons and the used assumption of the same transport physics, positrons will only contribute marginally to the diffuse gamma-ray emission.



\subsection{Gamma-ray emission}
\label{sc:GammaRays}

We distinguish between the local gamma-ray emissivity and their line-of-sight integrals, the diffuse gamma-ray emission (DGE).

\subsubsection{Local gamma-ray emissivity}
\label{sc:GammaRaysEmissivity}

\begin{figure}
	\setlength{\unitlength}{0.001\textwidth}
	\begin{picture}(1000,619)(-70,-70)
	\put(420,-40){$E$ [GeV]}
	\put(-40,180){\rotatebox{90}{$E^2 F$ [GeV/m$^3$ s sr]}}
	\includegraphics[width=930\unitlength]{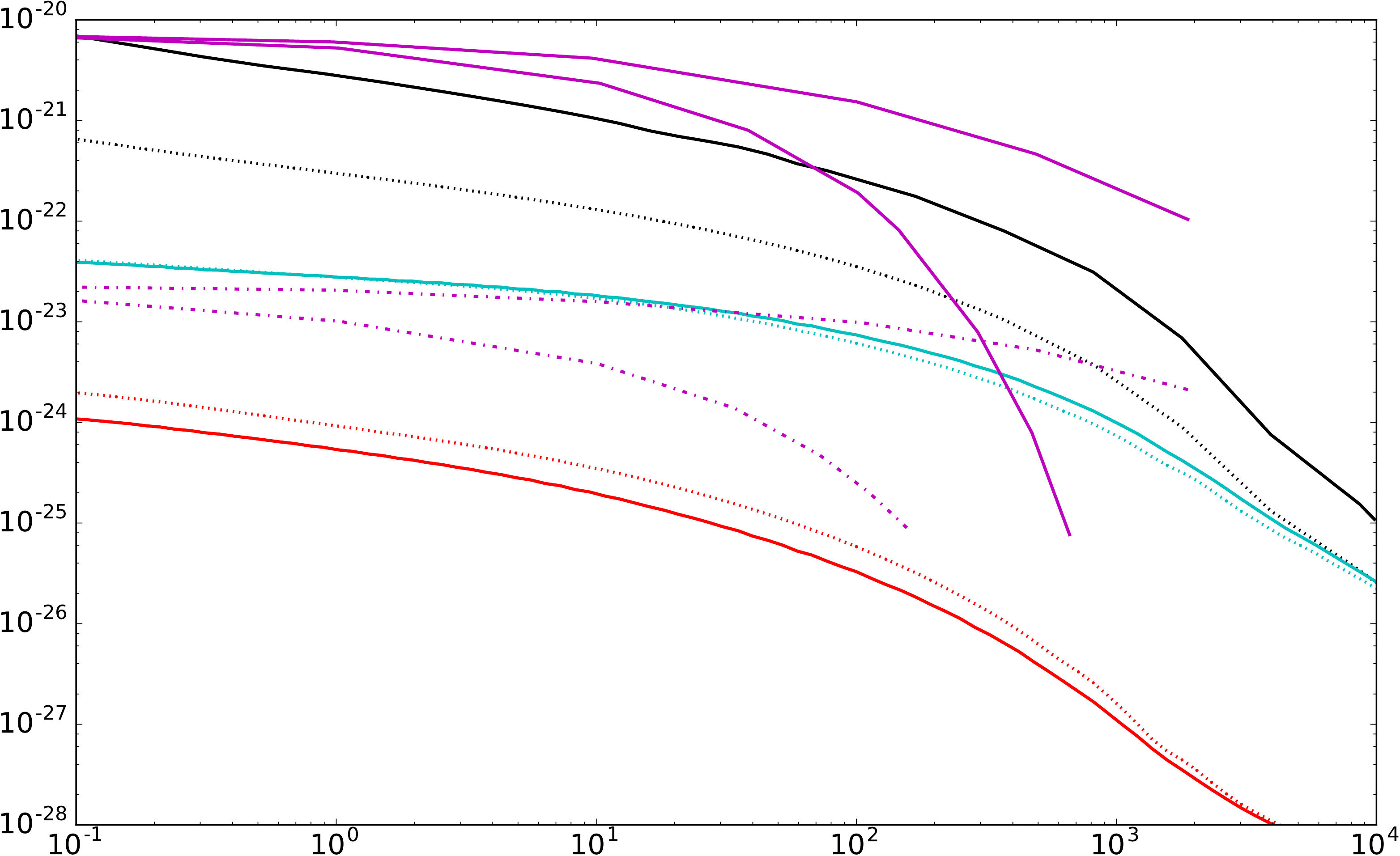}
	\end{picture}
	\caption{Local IC gamma-ray emissivities for both ISRF models 
		in the Galactic Center (black), at Earth (cyan) and at a halo position ($(10,10,2)$ kpc, red). Solid lines are for the emissivities using the \ISRFPicard model and dotted lines for using the \ISRFGalprop model. Magenta and green 
		shows the IC gamma-ray emissivities from \citep[Figure 13]{2017MNRAS.470.2539P} for
		comparison - Green shows the Galactic Center, Magenta Earth with
		electron-cut-off energies $E_b = 0.01\,$TeV (dashed) and $E_b = 1\,$TeV (dash-dotted).
	}
	\label{fig:Emissivities}
\end{figure}

The spatial and spectral differences in the two ISRF models lead to different local IC emissivities, i.e. the observed higher energy loss rates in the Galactic Center (see section \ref{sc:ApplicationEnergyLossRates}) are also reflected as correspondingly higher IC emissivities there, despite the significant reduction of the electron flux.
Figure \ref{fig:Emissivities} shows the spectrum for the
IC-related gamma-ray emissivity at the Galactic Center, Earth and a halo position at [10,10,2] kpc
for our models and for comparison also the IC emissivities from \citep[Figure 13]{2017MNRAS.470.2539P}.
The spectra following from the different ISRF models are similar in shape, but the model using the new ISRF leads to higher emissivities in the Galactic Center by up to an order of magnitude. At Earth the IC emissivities
are comparable and at the halo position the new ISRF model decreases the IC 
emissivity by approximately a factor of two.

\begin{figure}
	\setlength{\unitlength}{0.001\textwidth}
	\begin{picture}(1000,755)(-70,-70)
	\put(420,-40){$R$ [kpc]}
	\put(-40,230){\rotatebox{90}{$E^2 F$ [GeV/m$^3$ s sr]}}
	\includegraphics[width=930\unitlength]{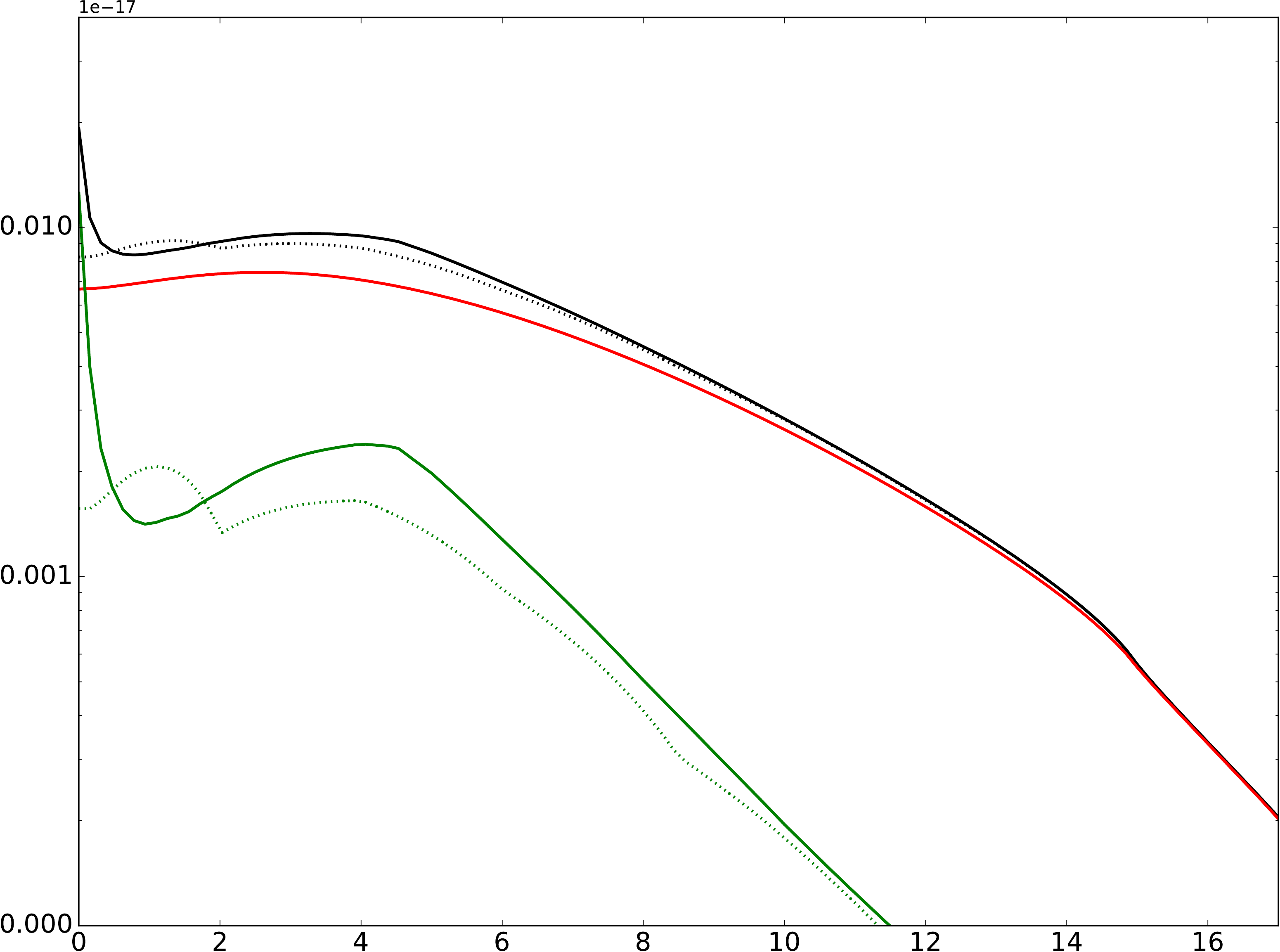}
	\end{picture}
	\caption{Radial profile of the IC (green) and total (black) gamma-rays emissivities at \SI{220}{GeV} 
		for both ISRF models. Results using the \ISRFPicard model are
		shown as solid lines whereas those for the \ISRFGalprop model are shown in dotted lines.
		Neutral pion decay emissivities are shown for comparison only (solid red).
		}
	\label{fig:EmissRadial}
\end{figure}

Figure \ref{fig:EmissRadial} then shows radial profiles for the IC, the neutral pion decay, and 
the total gamma-ray emissivities at \SI{220}{GeV}.

There is a notable increase in IC emission towards the Galactic Center, where IC becomes the dominant gamma-ray emission channel. Between \SI{100}{GeV} and \SI{1}{TeV} this characteristics is seen in a 150~pc-scale region around the Galactic Center, reaching the present spatial resolution limit in our simulations.

This IC emissivity increase in sub-kpc scales motivates further studies at
higher resolution (see section \ref{sc:Discussion}).

In contrast to \cite{2017MNRAS.470.2539P}, where a spatially constant CR electron spectrum has been assumed, our results show notably different energy dependencies of the gamma-ray emissivities at different galactocentric radii.


The highest gamma-ray emissivities outside of the Galactic Center are found at a distance of
$\sim$ \SI{5}{kpc} from the Galactic Center, coinciding with the peak in the chosen cosmic-ray source-distribution
\citep{2004A&A...422..545Y}.

Increased IC emissivity with respect to the \ISRFGalprop model
are found between \SI{2}{kpc} up to approximately \SI{10}{kpc} for gamma-ray energies higher than \SI{10}{GeV}.
In particular, the total emissivity increases considerably between
\SI{3}{kpc} and \SI{6}{kpc} and energies above \SI{100}{GeV} for the PICARD ISRF model.
The increase is a net effect of the used CR source distribution and the increased ISRF intensities.
A net increase towards the Galactic Center (for the inner \SI{0.1}{kpc} to \SI{0.3}{kpc} depending on the gamma-ray energy) is visible at all energies.

\subsubsection{Relative contributions to the DGE}
\label{sc:GammaRadiativeLosses}

In this section we analyse the contributions of the individual radiative processes to the
line-of-sight integrated total DGE.
The contributions are studied in the following directions: Galactic Center ($\sqrt{l^2 + b^2} < 3^{\circ}$), Galactic Plane
($|b| \le 5^\circ, |l| \le 180^\circ$), intermediate latitudes ($30^\circ < |b| < 90^\circ$), and the outer Galaxy ($|b| > 8^\circ$ or $|l| > 100^\circ$).

\begin{figure}
	\setlength{\unitlength}{0.001\textwidth}
	\begin{picture}(1000,322)(-55,0)
	\put(60,290){$\sqrt{l^2 + b^2} < 3^{\circ}$}
	\includegraphics[width=930\unitlength]{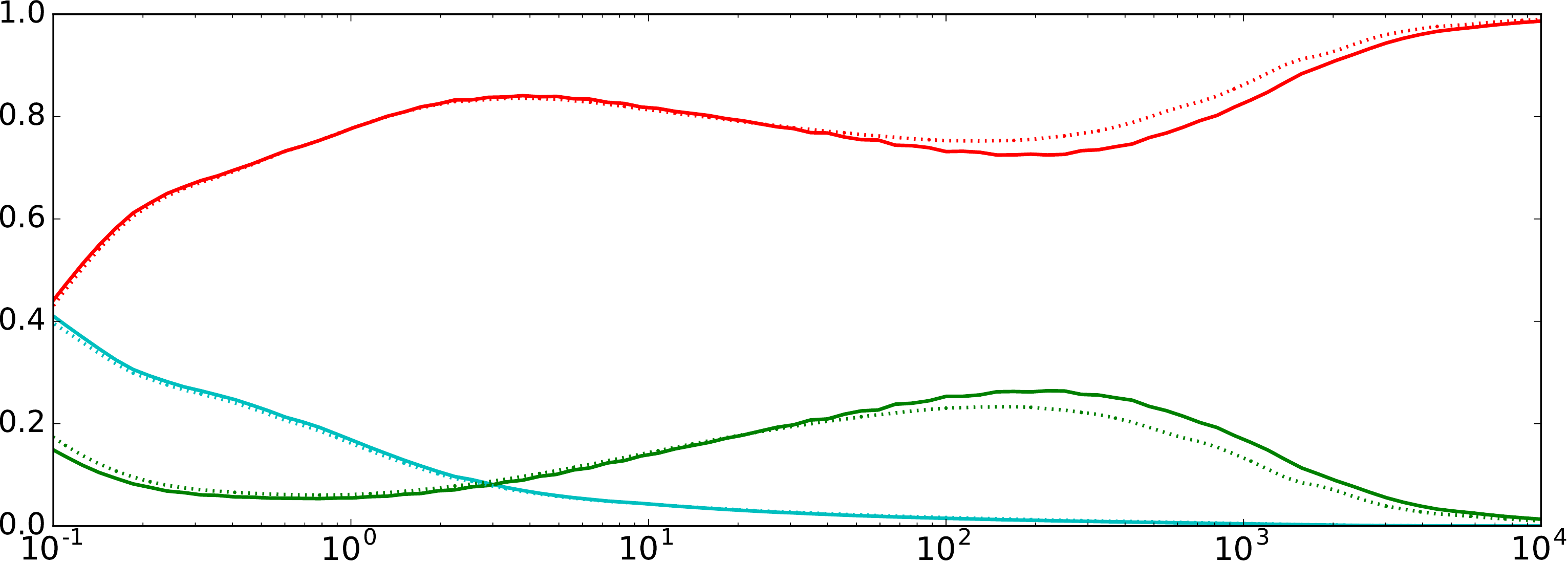}
	\put(-520,-35){$E$ [GeV]}
	\end{picture}
	\caption{Relative contributions of the different gamma-ray production channels to the DGE for the Galactic Center region. Neutral pion decay emission is in red, IC in green, Bremsstrahlung cyan. Dotted line for \ISRFGalprop model, solid lines for the \ISRFPicard model.
	}
	\label{fig:GammaRayContributions}
\end{figure}

Figure \ref{fig:GammaRayContributions} shows the contributions of the 
individual gamma-ray production channels for the Galactic Center region.
This is representative for all studied regions.
Neutral pion decay is the dominant gamma-ray production channel in all cases, with considerable contribution of the IC channel to the total gamma-ray emission at energies around 200~GeV.
The \ISRFPicard model increases the IC gamma-ray flux considerably between \SI{100}{GeV} and \SI{10}{TeV}.
IC related gamma-ray flux is maximum at \SI{220}{GeV} for the Galactic Center region and the Galactic Plane.

At this energy IC contributes  about $26\,\%$ towards the Galactic Center,
$25\,\%$ in the Galactic Plane, and $32\,\%$ at intermediate latitudes and the outer Galaxy to the total gamma-ray flux.
Lower IC losses and corresponding lower gamma-ray fluxes at higher energies relate to the transition into the Klein-Nishina-Regime.

\subsubsection{Gamma-ray spectra}
\label{sc:GammaRaysSpectra}

\begin{figure}
	\setlength{\unitlength}{0.001\textwidth}
	\begin{picture}(1000,325)(-55,0)
	\put(770,330){$r < 3^{\circ}$}
	\includegraphics[width=930\unitlength]{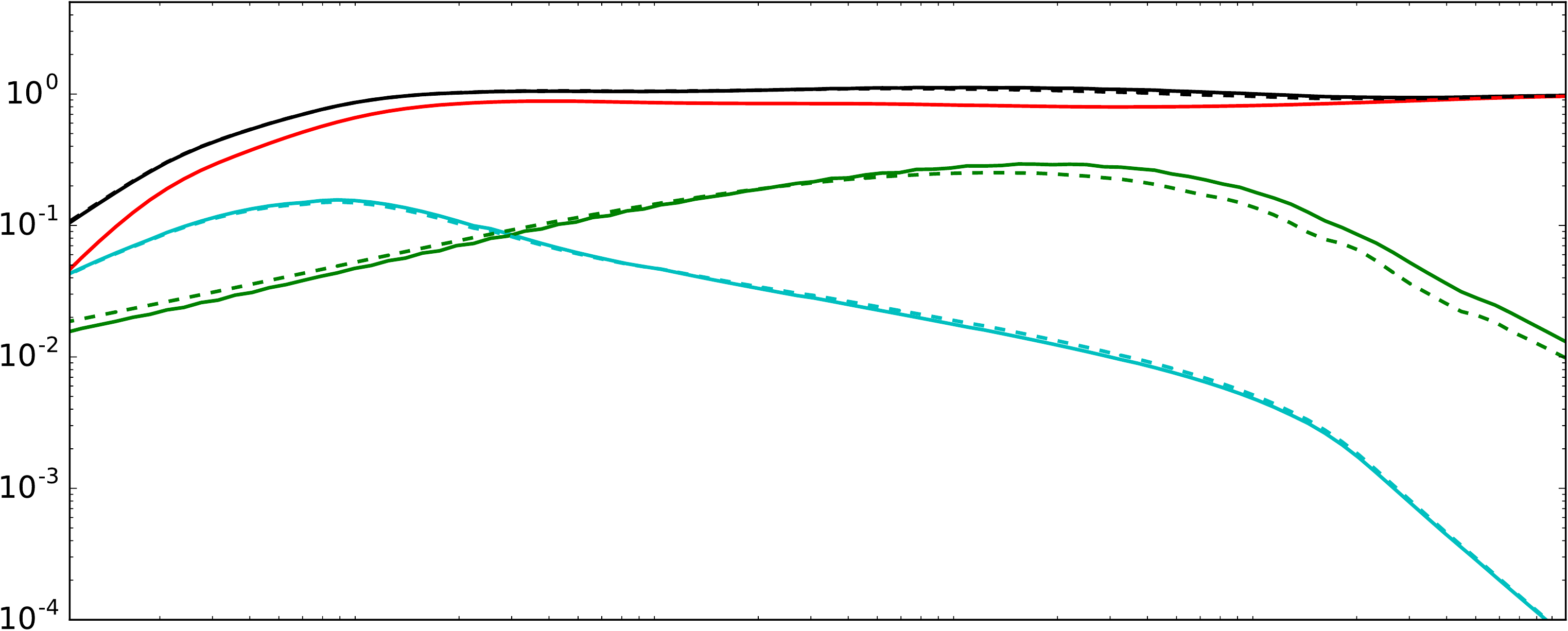}
	\end{picture}
	\begin{picture}(1000,375)(-55,0)
	\put(770,330){$|b| \le 5^{\circ}$}
	\includegraphics[width=930\unitlength]{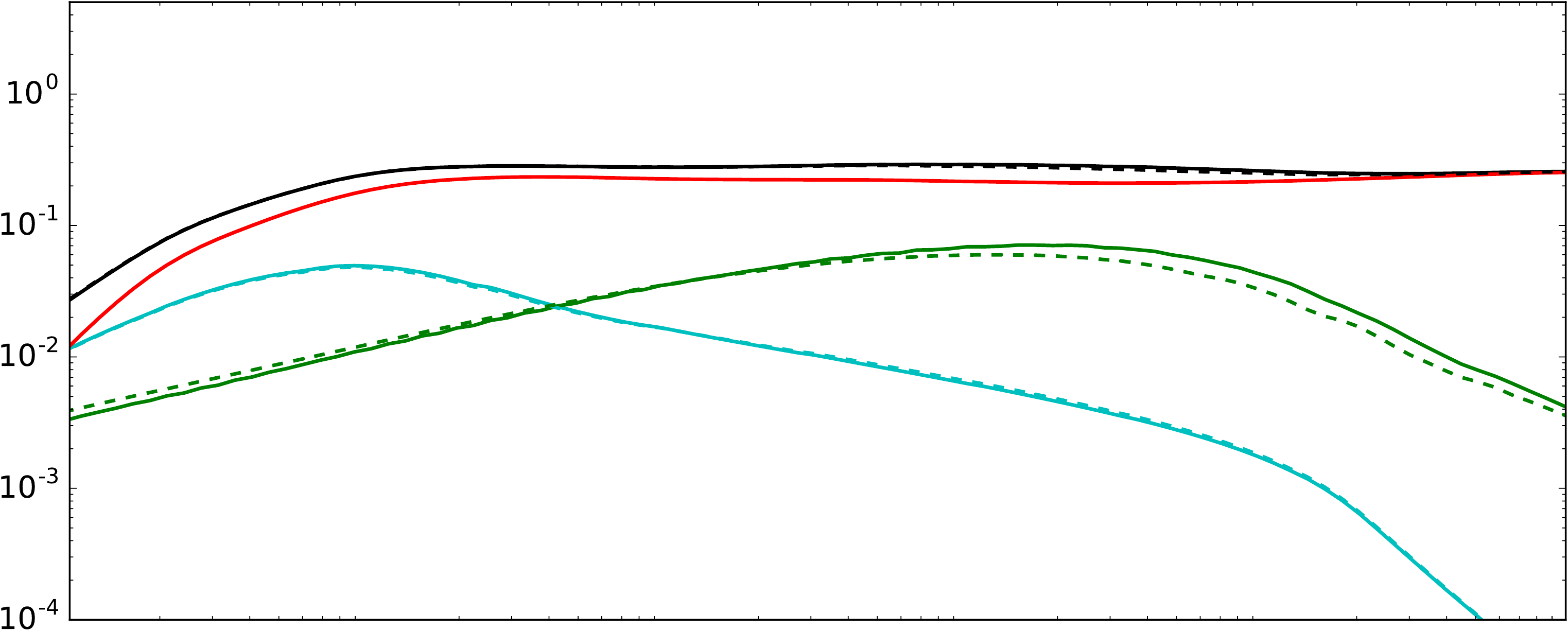}
	\end{picture}
	\begin{picture}(1000,375)(-55,0)
	\put(720,330){$10^{\circ} \le |b| \le 20^{\circ}$}
	\put(-35,200){\rotatebox{90}{$E^2.7 F$ [GeV$^{1.7}$ / m$^2$\,s\,sr]}}
	\includegraphics[width=930\unitlength]{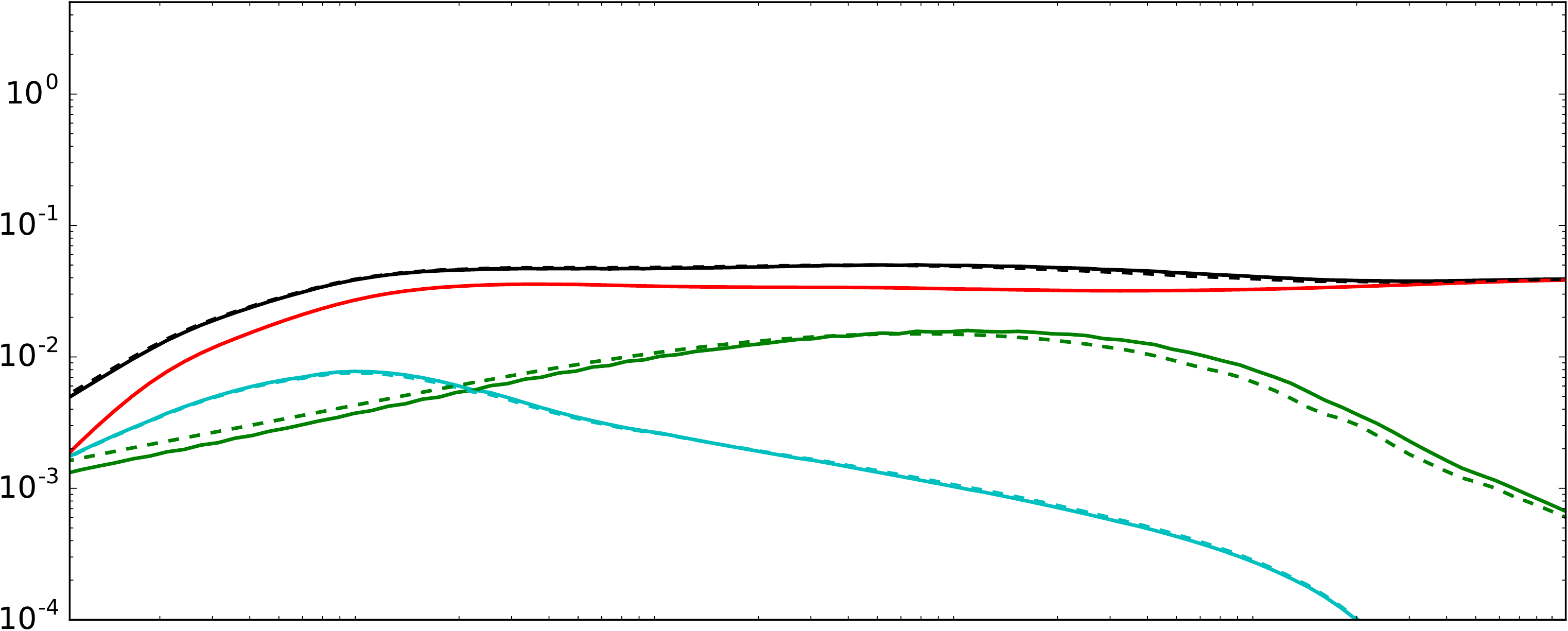}
	\end{picture}
	\begin{picture}(1000,395)(-55,0)
	\put(660,330){$|b| > 8^{\circ}$ or $|l| > 100^{\circ}$ }
	\put(420,-35){$E$ [GeV]}
	\includegraphics[width=944\unitlength]{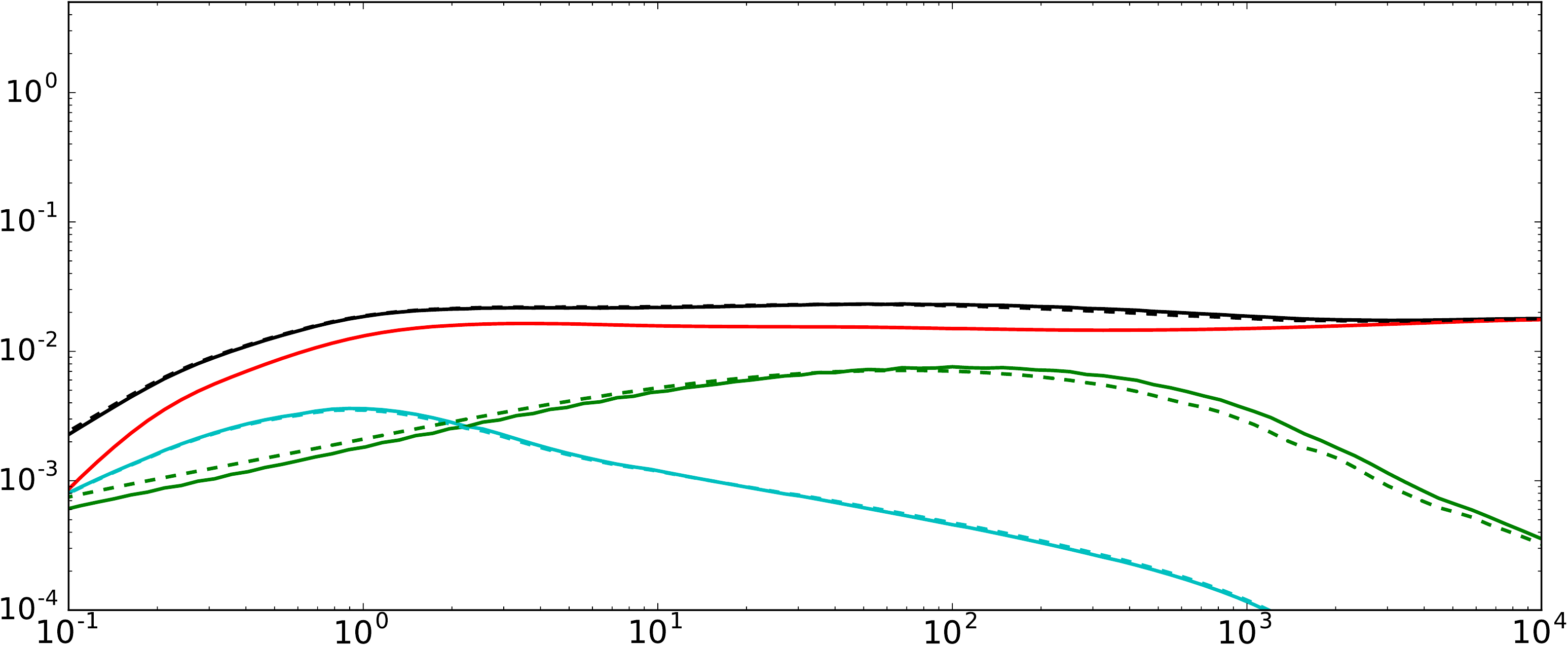}
	\end{picture}
	\caption{Gamma-ray emission spectra for different regions on the Sky. From top to bottom
		results are shown for: the Galactic Center, the Galactic Plane, intermediate latitudes, and the outer Galaxy.
		The specific extent of the regions is given in the respective sub-figure.
		Note that we use the same extent for the $y$-axis in all plots.
		Black is for total DGE, red is for $\pi^0$-decay, green for IC and cyan for Bremsstrahlung. Solid line for the (\textsc{Picard} ISRF), dashed line for the
		(\textsc{Galprop} ISRF).
	}
	\label{fig:GammaRaySpectra}
\end{figure}

Figure \ref{fig:GammaRaySpectra} shows the gamma-ray emission spectra for the three 
gamma-ray production channels (neutral pion decay, Bremsstrahlung, IC) for the two ISRF models
at the regions described in section \ref{sc:GammaRadiativeLosses}.
The DGE spectra of the two different ISRF models generally agree with each other.
Owing to the increased IC emission related to the new ISRF, also the total DGE increases
between \SI{20}{GeV} and \SI{10}{TeV} by up to 5~\%. The slight
decrease in IC emissivity between \SI{1}{GeV} and \SI{10}{GeV} translates to a negligible decrease of the total DGE of less than 1~\%.
%
The impact of the new ISRF on the gamma-ray emission differs for the different regions, but is highest for the Galactic Center and the Galactic Plane.

The \ISRFPicard model increases the spectral index with respect to the \ISRFGalprop model on average by $0.01$ between \SI{500}{GeV} and \SI{10}{TeV} for all regions.
Between \SI{15}{GeV} and \SI{500}{GeV} the spectrum hardens by $~0.01$ for the Galactic Center and the Galactic Plane. At intermediate latitudes and the outer Galaxy the spectral hardening is on average $0.02$ for the same energy regime.
The new ISRF model has a rather small impact on the spectral index on large scales.

\subsubsection{Gamma-ray spatial profiles}
\label{sc:GammaLonLat}

\begin{figure}
	\setlength{\unitlength}{0.001\textwidth}
	\begin{picture}(1000,503)(-70,0)
	\put(-40,150){\rotatebox{90}{$F_{I}$ [m$^{-2}$ s$^{-1}$ sr$^{-1}$]}}
	\includegraphics[width=930\unitlength]{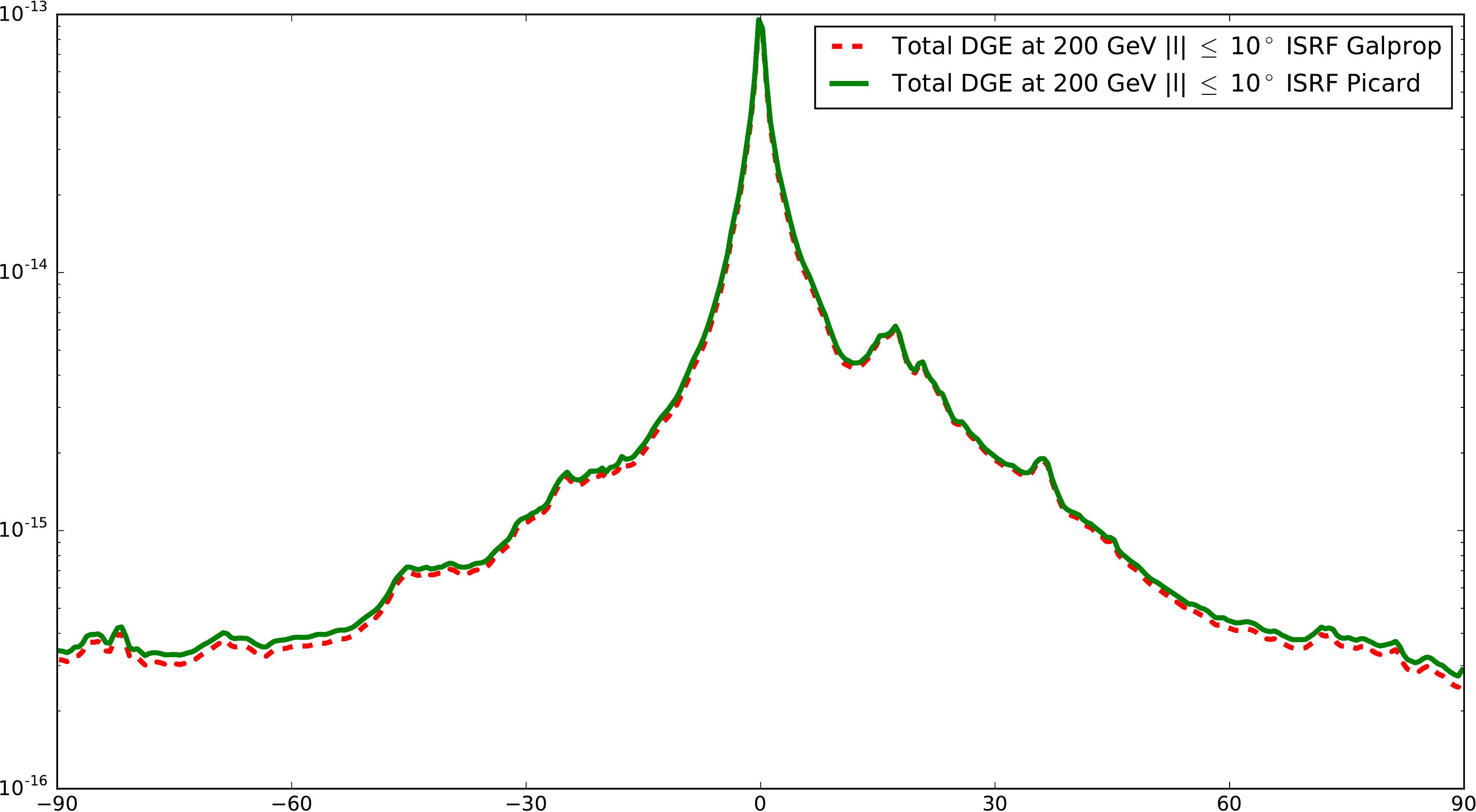}
	\end{picture}
	~\\
	\begin{picture}(1000,573)(-70,-70)
	\put(420,-40){$b$ [degree]}
	\put(-40,170){\rotatebox{90}{Residual (B-A)}}
	\includegraphics[width=930\unitlength]{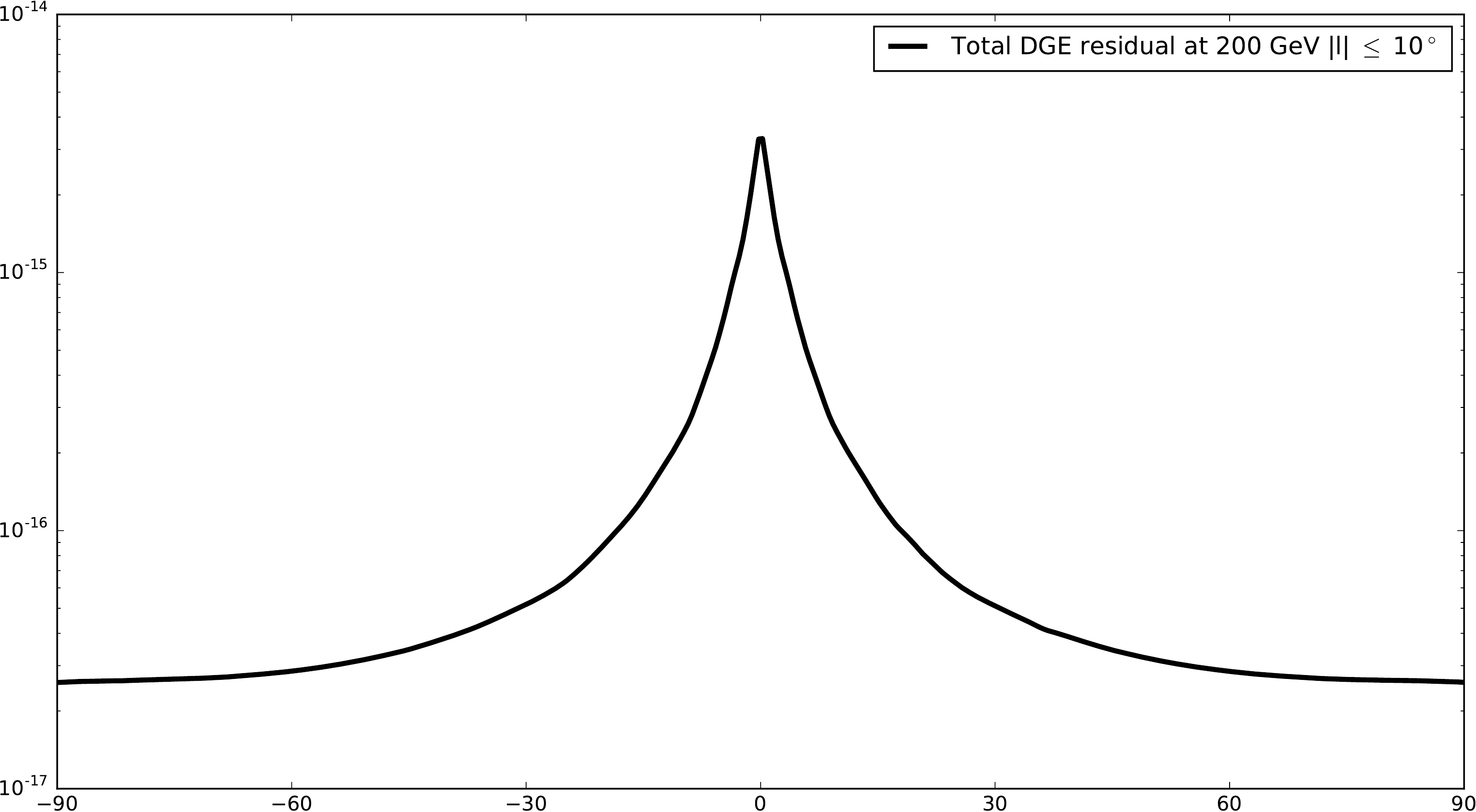}
	\end{picture}
	\caption{Latitude profile of the total DGE fluxes for \SI{220}{GeV} from $-10^\circ$ to $+10^\circ$ longitude for the two ISRF models (top) and residual between the total DGE fluxes (bottom). For the top
		Figure: Green solid line for the \ISRFPicard model, red dashed for the \ISRFGalprop model. For the residuals: A is the \ISRFGalprop model, B the \ISRFPicard model.}
	\label{fig:GammaRayLatProfiles}
\end{figure}

\begin{figure}
	\setlength{\unitlength}{0.001\textwidth}
	\begin{picture}(1000,503)(-70,0)
	\put(-40,150){\rotatebox{90}{$F_{I}$ [m$^{-2}$ s$^{-1}$ sr$^{-1}$]}}
	\includegraphics[width=930\unitlength]{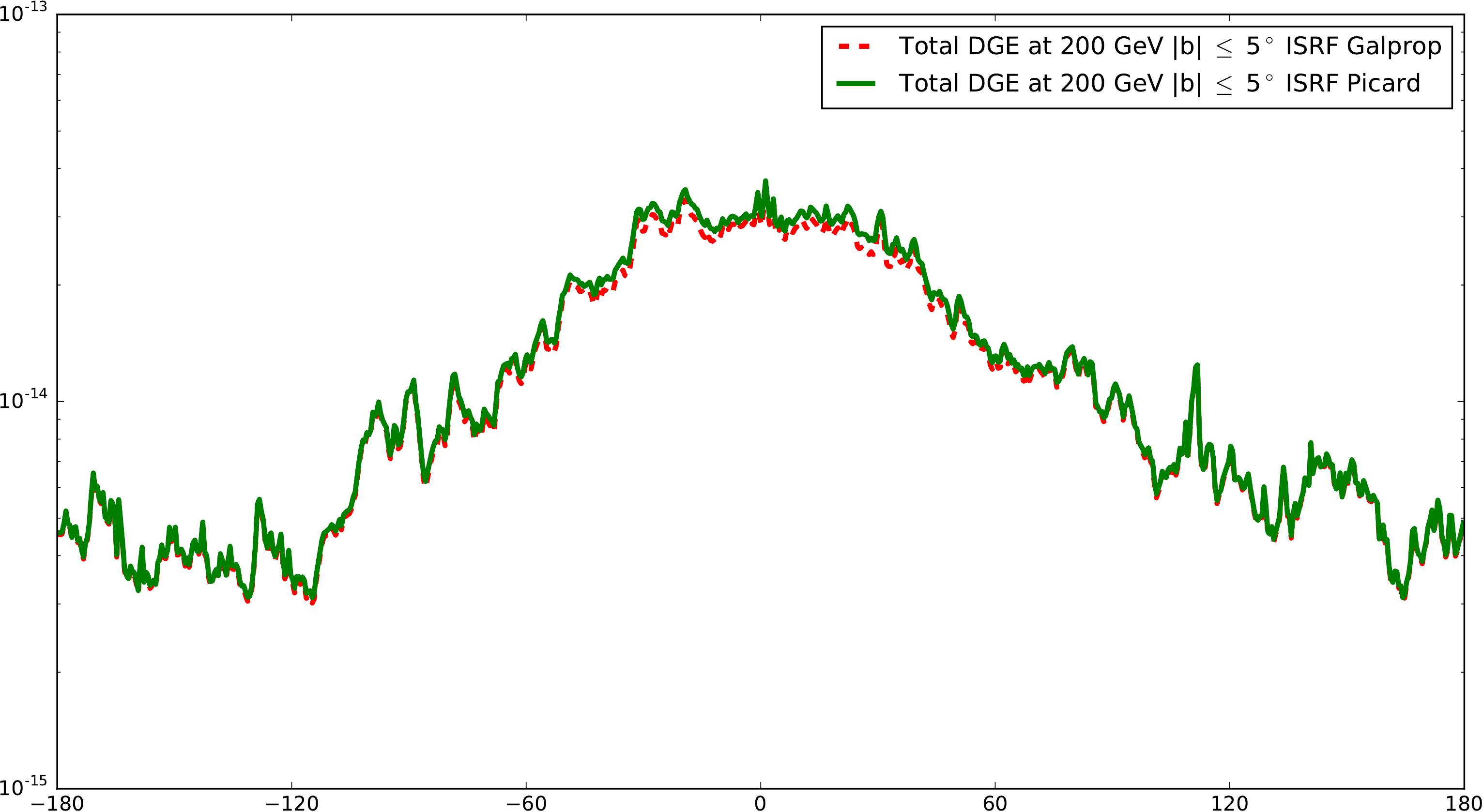}
	\end{picture}
	~\\
	\begin{picture}(1000,573)(-70,-70)
	\put(420,-40){$l$ [degree]}
	\put(-40,170){\rotatebox{90}{Residual (B-A)}}
	\includegraphics[width=930\unitlength]{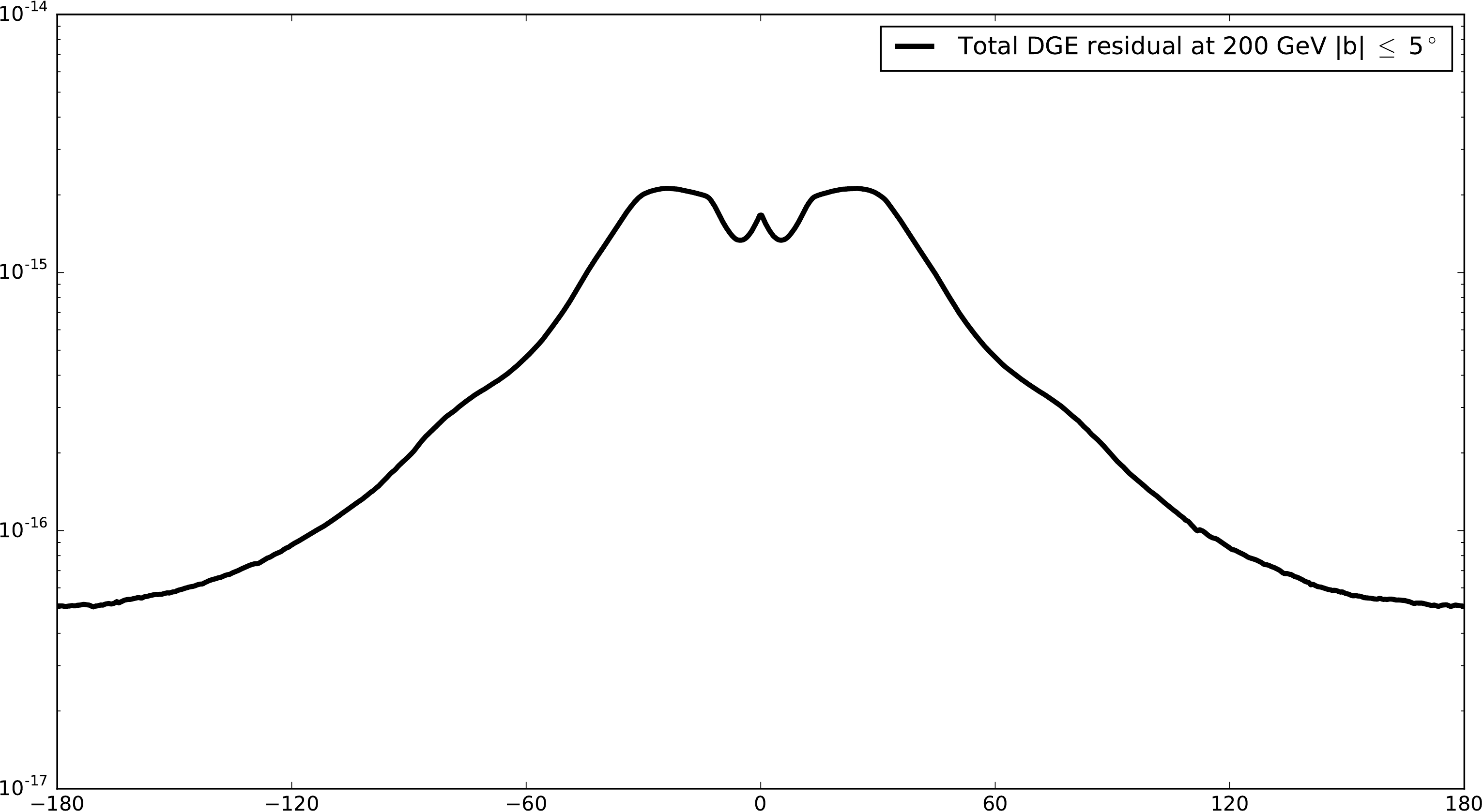}
	\end{picture}
	\caption{Same as Figure \ref{fig:GammaRayLatProfiles} but for the longitude profile over $|b| \leq 5^\circ$}
	\label{fig:GammaRayLonProfiles}
\end{figure}

We investigate the longitude and latitude profiles and corresponding residuals for the DGE at \SI{220}{GeV} where the new ISRF has the largest influence on the gamma-ray emission
(See Figure \ref{fig:GammaRayContributions}).
Latitude profiles are shown for $|l| \le 10^\circ$ in Figure \ref{fig:GammaRayLatProfiles} and longitude profiles for $|b| \le 5^\circ$ respectively Figure \ref{fig:GammaRayLonProfiles}.


The latitude and longitude profiles for both ISRF models agree well with each other but the DGE for the  \ISRFPicard model shows an increase in intensity at those energies.
A notable increase in diffuse Galactic emission towards the Galactic Center is evident in both spatial profiles.
The maximum DGE increases up to $4\%$ for the latitude profile at
\SI{220}{GeV}.
Considering the present resolution of the computations near the Galactic Center, it is expected that this difference becomes more pronounced at smaller scales.



In the longitude profile, the two maxima in the residual around $\pm 25^\circ$ are the consequence of the IC emissivity increase in the region between \SI{3}{kpc} and \SI{6}kpc from the Galactic Center (see Figure \ref{fig:EmissRadial}). 
The central peak in the residual reflect the higher IC emissivities at the innermost few hundred parsecs of the Galaxy.
%

\begin{figure}
	\setlength{\unitlength}{0.001\textwidth}
	\begin{picture}(1000,573)(-70,-70)
	\put(420,-40){$b$ [degree]}
	\put(-40,210){\rotatebox{90}{(A-B)/A}}
	\includegraphics[width=930\unitlength]{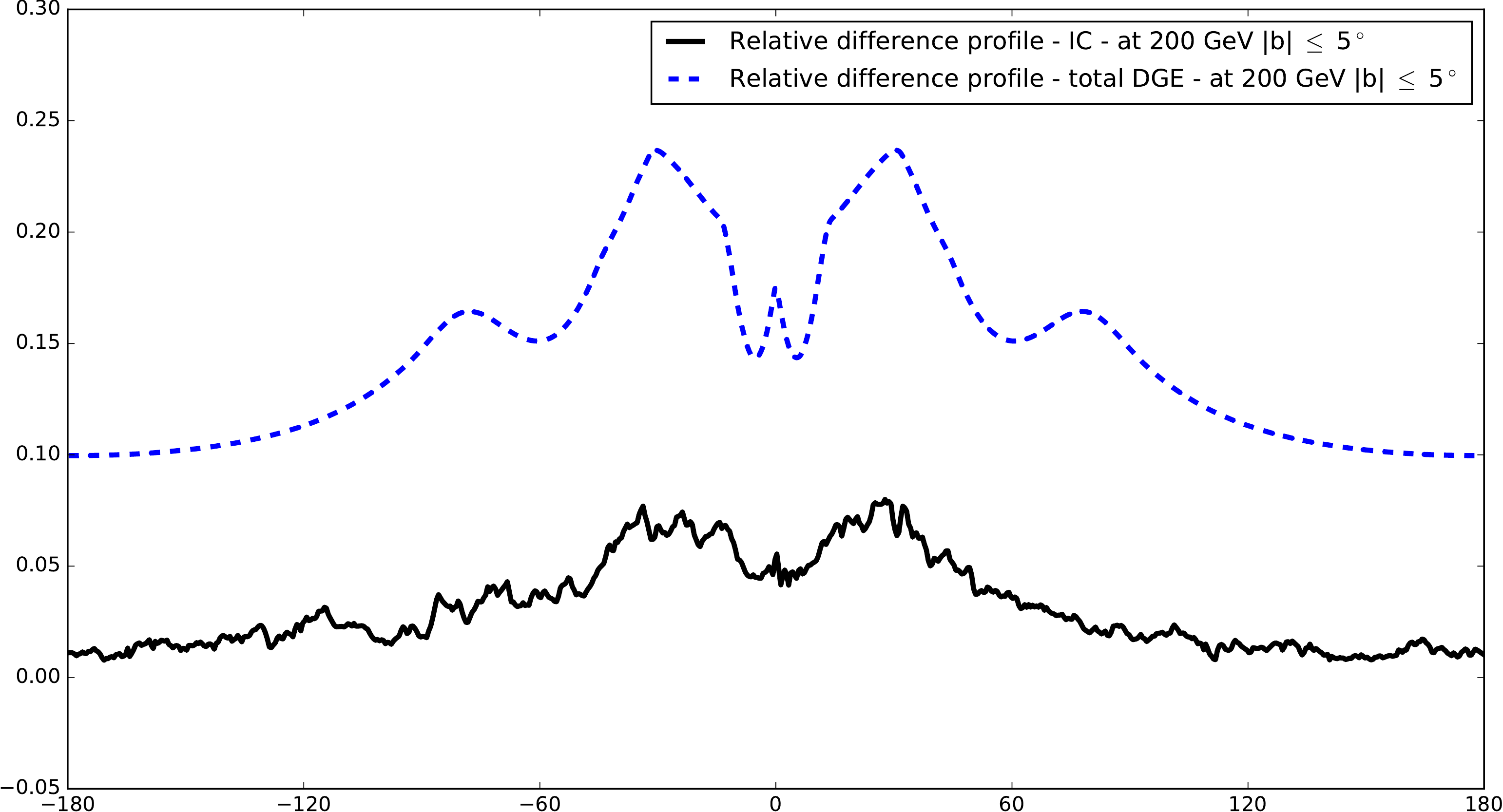}
	\end{picture}
	\caption{Longitude profile for the relative differences (A-B)/A in the total DGE fluxes (black, solid) and IC fluxes (blue, dashed) at \SI{220}{GeV} integrated over 
		$-5^\circ$ to $+5^\circ$ latitude for the two ISRF models.}
	\label{fig:GammaRayRelLonProfile}
\end{figure}

Figure \ref{fig:GammaRayRelLonProfile} shows the relative changes of the longitude profile ($|b| \lt 5^\circ$) for the IC and the total DGE fluxes between the two ISRF models.
The IC emission 
is increased by 25~\% at $\pm 30^\circ$ from the Galactic Center. The emissivity increase for the inner $0.1$-\SI{0.3}{kpc} of
the Galactic Center relate to an increase of up to $16\,\%$ in the IC flux of the longitude 
profile. Now for the total DGE, these increases account for changes of $2\%$-$5\%$ in
the total DGE profile, being especially interesting for the regions
from $-90^\circ$ to $60^\circ$.

\subsubsection{Gamma-ray skymaps}

\begin{figure}
	\begin{picture}(300,250)(0,0)
		\put(-40,80){\rotatebox{90}{IC emission - $(A-B)/A$}}
		\includegraphics[width=\textwidth]{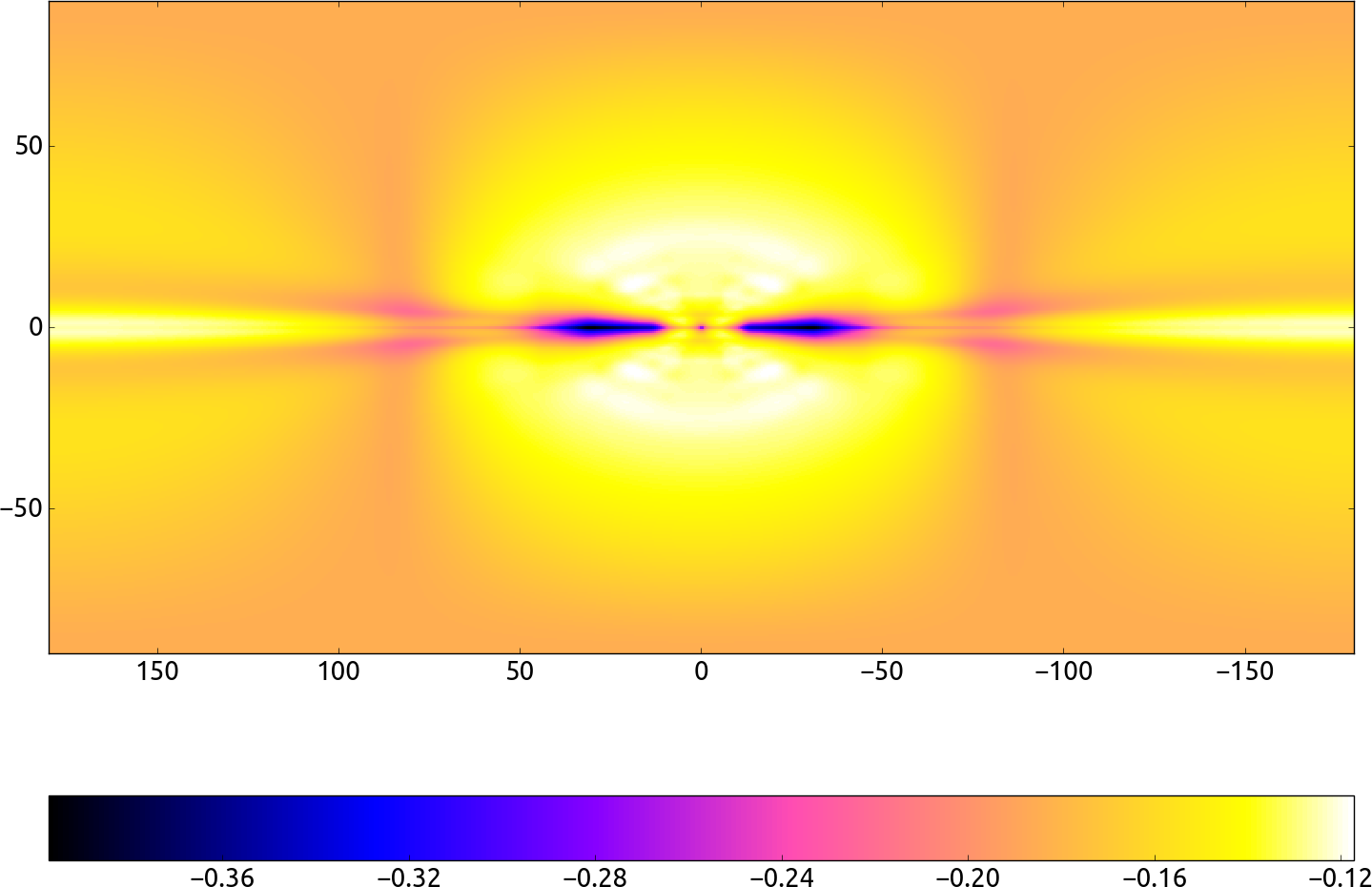}
	\end{picture}
	\begin{picture}(300,270)(0,0)
		\put(-40,90){\rotatebox{90}{Total DGE Residual - $(A-B)$}}
		\includegraphics[width=\textwidth]{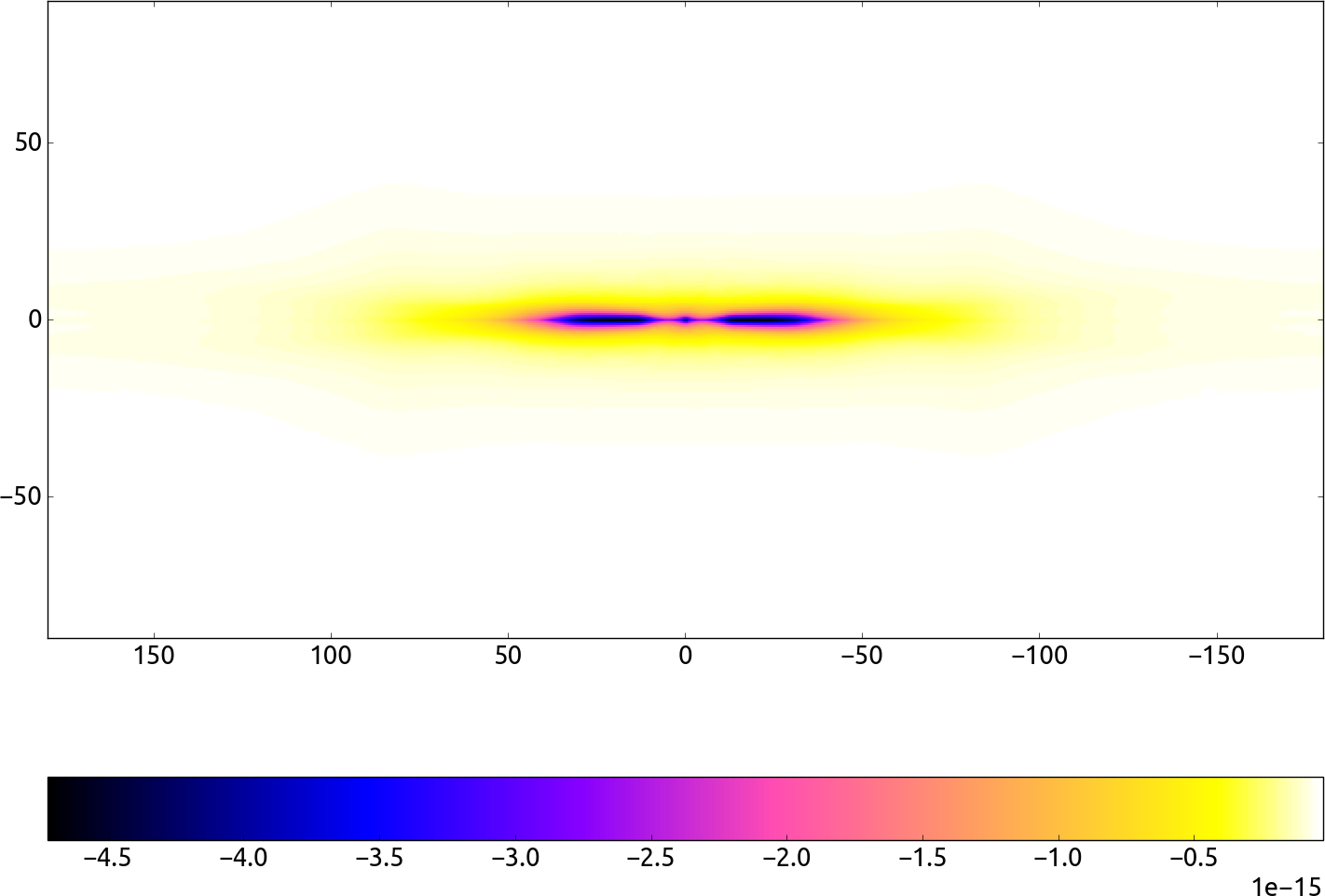}
	\end{picture}
	\caption{Relative differences in the IC emission (top) at \SI{220}{GeV} and residual for the total DGE emission (bottom) between the two ISRF models. Positive values indicate a
		surplus in gamma-ray emission for the \ISRFGalprop model, negative values for the
		\ISRFPicard model. We use an unbalanced colormap to increase the contrast.}
	\label{fig:GammaRayResidualMaps}
\end{figure}

In order to illustrate differences between gamma predictions in both ISRFs 
Figure \ref{fig:GammaRayResidualMaps} shows residual maps of the IC and the total DGE
emission at \SI{220}{GeV}, respectively.
The predictions differe distinctively in three different regions:
between $-45^\circ$ and $-10^\circ$ in longitude, between
$10^\circ$ and $45^\circ$ in longitude, and in the Galactic Center for $r \leq 2^\circ$. The former two are consequence of the chosen CR source distribution, and compares well with the longitude profile in Figure \ref{fig:GammaRayLonProfiles}.
At $l = \pm 30^\circ$ these structures extent to up to $8^\circ$ in latitude.
%
%
Evidently, the \ISRFPicard model further amplifies features of the underlying axisymmetric source model.
%

\newcommand{\newISRF}{PICARD ISRF }

\section{Discussion}
\label{sc:Discussion}

In this paper we have studied effects resulting from the use of the new ISRF model \citep{2017MNRAS.470.2539P} on the diffuse gamma-ray emission in a regime from GeV to \SI{10}{TeV}.
Compared to the widely used \ISRFGalprop model, the \ISRFPicard model 
introduces three notable changes with consequences for predictions of diffuse gamma-ray emission:

\begin{itemize}
	\item Order-of-magnitude increased intensity towards the Galactic Center
	\item Generally increased intensity throughout the dust-dominated regime
	\item Reduced intensity at shorter wavelengths in the starlight regime
\end{itemize}

We confirm the findings of \citet{2017MNRAS.470.2539P} in a cosmic-ray propagation model where we solve the transport equation for an axisymmetric CR source model and show quantitatively 
the resulting consequences for the DGE.

The local gamma-ray emissivities in the Milky-Way show a considerable increase for all energies in the Galactic Center, that rapidly decreases with increasing Galactocentric
distance. For the inner 0.1-0.3~kpc we find an order of magnitude increase in IC emissivity
(the size of this region decreases with gamma-ray energy). Thus, at Galactic Center IC becomes the dominant production channel for diffuse gamma-ray emission.
Those findings motivate further studies on small scales where the 
localized emissivity increases
are expected to 
become even more pronounced
when using sub-kpc resolution in conjunction with the \ISRFPicard model.

The emissivities at Earth are comparable while showing a mild decrease 
in the Galactic halo.

Despite a reduction in the electron flux due to the higher intensity of the \ISRFPicard at the Galactic Center, the IC emissivity is increased by an order-of-magnitude.
In contrast, higher ISRF intensity in the dust-dominated regime does not affect the energy loss rates and electron spectra at Earth (see Figure \ref{fig:LeptonSpectra}).

Only recently, the \ISRFGalprop has been revised \citep{0004-637X-846-1-67}, also leading to increased gamma-ray emission.
Likewise, consideration of 3D structure in our Milky-way is advocated regarding the construction of the ISRF \citep{0004-637X-846-1-67}, the gas distribution \citep{johannesson2017interstellar}, and principal execution of galactic propagation \citep{porter2017high} as has already been extensively discussed in earlier works using the dynamic Milky-way model \citep{2016ApJ...826...47B}, DRAGON \citep{2013PhRvL.111b1102G}, and PICARD \citep{2015APh....64...18W,2015APh....70...39K}. In  light of these recent developments we consider it desirable to compare how this alters results and conclusions from earlier numerical models of cosmic ray propagation and associated gamma-ray emission, most particular with emphasis on the IC-induced gamma-ray emission.
As the importance of the transition from axisymmetric models for CR source distribution to more realistic models, e.g. including spiral-arms, has been already discussed previously \citep{2015APh....64...18W,2013PhRvL.111b1102G,2016ApJ...826...47B} we can compare now realisations of different ISRFs and gamma-ray predictions originating
therefrom. Whereas the \ISRFPicard \citep{2017MNRAS.470.2539P} was build upon axial-symmetry, it is constructed on logarithmic binning, allowing for very high resolution towards the Galactic Center: the ISRF portrayed in \citep{0004-637X-846-1-67} rests on a 3D-Cartesian Grid with lower resolution in the Galactic Center/Bulge/Disk.
This becomes relevant for the intensity increase of the gamma-ray emissivity in the inner $300$ pc of our Galaxy, corresponding to the inner $2^\circ$, well
located in the region, where a noticeable Galactic Center excess has been reported by Fermi-LAT \citep{2012ApJ...750....3A}.
By using our rather coarse Galaxy model we can only account for a flux increase up to 5~\% at the Galactic Center, thus
the ISRF model alone lacks an amplification effect that is sufficient large to account for
the inferred gamma-ray intensity increase towards the Galactic Center.
The ISRF model consequently needs
to be discussed in conjunction with more accurate prescriptions for the used CR source component, as e.g. discussed in\citep{2015APh....64...18W,0004-637X-846-1-67}.

Given that observed gamma-ray excess emission towards the Galactic Center \citep{2012ApJ...750....3A} is frequently discussed in relation to dark matter in the inner Galaxy \citep{2009arXiv0910.2998G,HOOPER2011412,HOOPER2013118,2016PDU....12....1D,2012PhRvD..86h3511A,2015PhRvD..91f3003C,PhysRevD.89.049901}, the application of the \ISRFPicard \citep{2017MNRAS.470.2539P} in numerical frameworks for Galactic CR propagation, exercised here with PICARD, offers an interesting alternative alongside proposed alterations of the source distribution \citep{2015JCAP...12..005C,PhysRevD.84.123005}, or different prescriptions of Galactic diffusion in a propagation model like DRAGON (\citep{1475-7516-2015-12-056,2008JCAP...10..018E} in explaining the Galactic Center excess. Also the effect of recent injections of high-energy cosmic rays in the Galactic Center region as e.g. suggested by
\cite{2014PhRvD..90b3015C,2015JCAP...12..005C} would additionally enhance the respective gamma-ray yields when using the \ISRFPicard model. 
While different attempts to explain the Galactic Center excess have been made, distinguishing the different components by observations remains difficult. While we could show, that 
the \newISRF model leads to a distinctive increase in the Galactic Center and towards the Galactic Plane, it does not provide a method for observational distinction between the different emission scenarios.

It is necessary to distinguish a strongly-peaked excess in the  Galactic Center  \citep{PhysRevD.84.123005}, relating to possible dark matter processes,
from an extended excess like reported in \citep{0067-0049-223-2-26}.
\citet{1475-7516-2015-12-056}  find for the later that models with conventional CR physics show at least as good a fit to such an excess as DM related ones \citep{1475-7516-2015-12-056}.
The \newISRF model can account for both excess types, the very localised and an extended component, but has limitations:
For the increase in the Galactic Center it is important to mention that the 
used source distribution using pulsars \citep{2004A&A...422..545Y} has its largest uncertainties there. Since IC-related gamma-ray emission is influenced by the local ISRF intensities and the used
CR source distribution, reliable predictions in that region still remain difficult.
For a larger-scale component, the \newISRF model can account for an increase
in the higher GeV to TeV regime, well motivated by previous studies like \citep{2012ApJ...750....3A}, where they account for a 
possible large-scale component by applying scaling parameters for the ISRF.
By introducing further alterations regarding the assumed CR source distribution alongside choosing more adequate Galactic models \citep{2013arXiv1306.6850G,2015APh....70...39K,POS2017_279-1}
one might gain sufficient new/underexplored degrees of freedom to provide explanations for the Galactic Center excess problem within conventional CR propagation scenarios without hypothesizing contributions from dark matter annihilation.

Lastly, $\gamma\gamma$-attenuation becomes a relevant competing process at energies above \SI{10}{TeV}, as most recently studied in \citep{2017MNRAS.470.2539P}.
Here, we restrict ourself to energies where attenuation did not affect the predictions beyond  percent-level. We refrain from expanding our energy range presently owing to the transition from the diffusion-dominated regime at GeV energies to the source-dominated regime at TeV energies and associated uncertainties in 3D-formulation of the CR source distribution. Numerically efficient 3D propagation models like PICARD now offer the possibility to study 3D source distribution at deca-parsec resolution, and therefore hold promise to deal with adequate propagation physics near CR sources. Recent reports on a CR Pevatron in the Galactic Center by the H.E.S.S. collaboration \citep{2016Natur.531..476H} strongly supports altering or complementing presently used (naive) CR source distributions for application in advanced Galactic CR propagation frameworks. This will be thoroughly investigated in a forthcoming study.

\section{Acknowledgements}

The financial support from the Austrian Science Fund (FWF) project number
\textit{I1345}, in collaboration with the French National Research Agency (ANR),
project ID \textit{ANR-13-IS05-0001} is acknowledged.
The computational results presented have been achieved (in part) using the HPC infrastructure LEO and MACH of the University of Innsbruck.
Cristina C Popescu acknowledges support from the Leverhulme Trust Research Project Grant RPG-2013-418.

\newpage



\bibliography{Paper}

\end{document}